\documentclass[aps,pra,twocolumn,superscriptaddress,amsmath,amssymb,floatfix]{revtex4-2}
\usepackage[normalem]{ulem}
\usepackage{graphicx}
\usepackage{dcolumn}
\usepackage{bm}
\usepackage{physics}
\usepackage{xcolor}
\usepackage{mleftright}
\usepackage{float}
\usepackage[colorlinks]{hyperref}
\hypersetup{plainpages=true,breaklinks=true,hypertexnames=false,pageanchor=true,
  colorlinks=true,linkcolor={blue},citecolor={red},urlcolor={blue},anchorcolor={black}}

\renewcommand{\eqref}[1]{\mbox{Eq.~(\ref{#1})}}
\newcommand{\eqrefs}[2]{\mbox{Eqs.~(\ref{#1},\,\ref{#2})}}

\newcommand{\figref}[1]{\mbox{Fig.~\ref{#1}}}
\newcommand{\secref}[1]{\mbox{Sec.~\ref{#1}}}
\newcommand{\appref}[1]{\mbox{Appendix~\ref{#1}}}

\begin{document}

\title{Finite-Temperature Superradiant Criticality\\and Thermal Emission in the Open Dicke Model}

\author{Gabriele Orlando}
\author{Samuel Napoli}
\author{Salvatore Savasta}
\author{Omar Di Stefano}
\affiliation{Dipartimento di Scienze Matematiche e Informatiche, Scienze Fisiche e Scienze della Terra, Universit\`a di Messina, I-98166 Messina, Italy}

\begin{abstract}
Near the superradiant phase transition of the Dicke model, the lowest-frequency excitation softens and its losses can no longer be treated as a weak perturbation. We develop a finite-temperature theory of this transition in the open system, with no restriction on the strength of the system-bath coupling. The cavity mode and the atomic ensemble are coupled to independent reservoirs, the first at zero temperature and the second at temperature $T$, and we assume that the spins remain thermalized with their reservoir. We find that the reservoirs do not shift the temperature-dependent critical coupling for a broad class of spectral densities, while the polariton becomes overdamped in a finite interval of couplings around it. As expected, raising the temperature moves the transition to stronger couplings, and in the superradiant phase we find a faster decoupling of light and matter. The theoretical framework developed here yields the reflection and thermal emission spectra for arbitrary parameters in both the normal and the superradiant phase, and provides a basis for comparison with experiments on systems displaying ground-state condensation. We also identify conditions under which the system behaves, at specific frequencies, as both a perfect absorber and a perfect emitter.
\end{abstract}

\maketitle

\section{Introduction}
\label{sec:intro}

The Dicke model describes a single bosonic mode coupled collectively to $N$ two-level systems~\cite{Dicke1954,Garraway2011,Kirton2019}. In the thermodynamic limit ($N\to \infty$) it displays a second-order equilibrium phase transition: above a critical coupling, the cavity field and the collective atomic polarization both become macroscopic in the absence of any external drive. This superradiant phase transition (SPT) was first established in thermal equilibrium by Hepp and Lieb and by Wang and Hioe~\cite{HeppLieb1973AnnPhys,WangHioe1973}, and the analysis was then extended to include the counter-rotating terms of the interaction~\cite{HeppLieb1973PRA,Hioe1973,Duncan1974}. Functional-integral methods later gave the partition function~\cite{PopovFedotov1982,
AparicioAlcalde2011} and the collective excitations of
the model~\cite{AparicioAlcalde2011}. At zero temperature the transition becomes a quantum phase transition~\cite{NatafCiuti2010,Sachdev2011}, characterized by the softening of the lower polariton~\cite{EmaryBrandes2003PRL,EmaryBrandes2003PRE} and by a divergence of the atom-field entanglement~\cite{Lambert2004,Lambert2005PRA}.

The experimental realization of the equilibrium SPT has been debated for a long time. No-go theorems show that the transition is forbidden for dipoles that interact with the electromagnetic field only through their electric polarization~\cite{Rzazewski1975,Knight1978,NatafCiuti2010,Viehmann2011,DeBernardis2018,Andolina2019,Andolina2022}, a conclusion that has recently been reexamined for many-dipole systems including dipole-dipole interactions~\cite{Lamberto2025PRR}.  Proposals to overcome these limits consider electrons with Rashba spin-orbit coupling~\cite{Nataf2019}, electrons or molecules coupled to spatially varying cavity fields~\cite{Andolina2020,Guerci2020,Mercurio2024}, and magnetic molecules coupled to the cavity through the Zeeman interaction~\cite{RomanRoche2021}. Recently, spectroscopic evidence for an equilibrium SPT was reported in ErFeO$_3$~\cite{Kim2025}, where Er$^{3+}$ spins interact with Fe$^{3+}$ magnons that play the role of photons, as predicted theoretically in Ref.~\cite{Bamba2022}. In this experiment the signatures of the transition are the softening of the lower polariton and a kink in the upper polariton, and the phase boundary is a line in the plane of temperature and magnetic field. Temperature is therefore a control parameter of the measurement and has to be included in its theoretical description. Effective Dicke Hamiltonians have also been proposed and realized in driven-dissipative systems, such as cold atoms in optical cavities~\cite{Dimer2007,Baumann2010,Nagy2010,Klinder2015,Zhiqiang2017,Bolian2025} and trapped ions~\cite{GenwayPRL2014, SafaviNaini2018,CohnNEWJPHYS2018,GambettaPRA2019}. Although these systems show strong analogies with the equilibrium SPT, their transitions are different in nature~\cite{Kirton2019}.

Theoretical work on dissipation in the Dicke model has mostly addressed these driven platforms. Master-equation and Keldysh treatments predict a critical point that depends on the cavity loss rate~\cite{DallaTorre2013,Kopylov2013,Brange2024}, while local atomic decay and dephasing can suppress or restore the transition~\cite{KirtonKeeling2017}. Mean-field equations become exact for such models in the thermodynamic limit~\cite{Mori2013,Carollo2021}, and their quantum correlations can be obtained from fluctuation operators~\cite{Boneberg2022}. The rotating-wave description of the reservoirs used in many of these works is appropriate in the frame imposed by the external drive, but it has no justification at equilibrium. In that case, the frequency of the lower polariton vanishes at the critical point, so that the ratio between the loss rate and the mode frequency grows without bound. Close enough to the transition, the system therefore enters first the ultrastrong and then the deep-strong regime of system-bath coupling~\cite{Lamberto2025}. In classical terms the lowest-frequency normal mode behaves as a damped harmonic oscillator in the overdamped regime~\cite{dekker1981classical}, in which the mode relaxes without oscillating. Perturbative descriptions of the reservoirs, such as standard master equations, break down in this regime.

Usual treatments of dissipation developed for ultrastrong light-matter coupling~\cite{FriskKockum2019,FornDiaz2019,Beaudoin2011,Ridolfo2012,Settineri2018,Mercurio2023} still assume a weak interaction with the environment, and the recent linear response theory of Ref.~\cite{RomanRoche2025} does not include the system-bath interaction in the Hamiltonian. These limits are removed in the zero-temperature theory of Ref.~\cite{Lamberto2025}, which is based on quantum Langevin equations that are not restricted by the smallness of the loss rates. That work shows that the critical point is not affected by reservoirs that interact with the system through a potential with a metastable minimum, and that the bath fields acquire macroscopic occupations in the superradiant phase, without changing the system condensates.

A finite-temperature counterpart of this description is still missing. Temperature changes the critical coupling and reduces the atomic polarization, which in turn enters both the collective light-matter coupling and the effective loss rate of the matter excitations. It also adds a new observable, namely the thermal emission of the system, which is strongly weighted toward low frequencies, where the lower polariton softens. The usual zero-temperature treatments of the Dicke model~\cite{EmaryBrandes2003PRL,EmaryBrandes2003PRE,Lambert2004} rely on a Holstein-Primakoff mapping within the sector of maximal total spin. A thermal state, however, has weight on all the total-spin sectors, so that a finite-temperature theory has to go beyond this restriction. The finite-temperature theories cited above~\cite{HeppLieb1973AnnPhys,WangHioe1973,PopovFedotov1982,AparicioAlcalde2011} treat the closed system. The same holds for the Majorana-fermion study of Alavirad and Lavasani, which does not assume total-spin conservation and provides an effective theory of the superradiant phase~\cite{Alavirad2019}. To our knowledge, finite-temperature analyses of spectral and input-output properties of the Dicke model either include damping phenomenologically or incorporate a thermal reservoir within a master-equation or Heisenberg-Langevin framework assuming weak dissipation, often addressing finite-size systems or the single-atom limit, see \textit{e.g.}~\cite{ImaiYamanaka2018,YamamotoTokura2025,Wang2026}.  Reference~\cite{DallaTorre2016} treats the thermodynamic limit without assuming total-spin conservation and describes a thermal bath acting on each atom with Lindblad operators. In that driven-dissipative setting the closed-model critical line is recovered only for vanishing decay rates, while for finite rates the critical coupling depends on both the cavity loss and the spin relaxation rate. All these approaches require loss rates much smaller than the system frequencies, a condition that fails near the critical point, where, as at zero temperature, the soft mode can become overdamped.

In this work we extend the theory of Ref.~\cite{Lamberto2025} to finite temperature. We consider a cavity mode and an atomic ensemble coupled to two independent bosonic reservoirs, with the photonic bath at zero temperature and the matter bath at temperature $T$, and we assume that the spins remain thermalized with their reservoir. We first determine the thermal mean-field state of the atoms, keeping all the total-spin sectors. The resulting closed-form critical line $g_c(T)$ coincides with the equilibrium result for the closed model including the counter-rotating terms~\cite{HeppLieb1973PRA,Hioe1973,Duncan1974}; it reduces to $\sqrt{\omega_a\omega_b}/2$ at $T=0$ (where $\omega_a$ and $\omega_b$ are the cavity and atomic frequencies) and grows as $\sqrt{T}$ at high temperature. The collective spin fluctuations about this state are described with the quantum central limit theorem~\cite{Goderis1989,Goderis1990,Benatti2015}. The reservoirs are modeled as independent oscillators with the same shifted coupling potential adopted in Ref.~\cite{Lamberto2025} (see also Refs.~\cite{CaldeiraLeggett1983,Ford1988}), and they are eliminated at the level of the Heisenberg equations of motion without invoking the Born-Markov or rotating-wave approximations. In this way we obtain quantum Langevin equations with a general memory kernel~\cite{GardinerZoller2004}. We specialize to ohmic baths~\cite{Leggett1987}, corresponding to frequency-independent damping rates, only after developing the general formalism.

Within this framework we obtain the following results. With the shifted coupling potential and for any frequency-dependent damping rate $\gamma(\omega)$ satisfying $\omega\gamma(\omega)\to 0$ as $\omega\to 0$, the reservoirs leave $g_c(T)$ unchanged, as found at zero temperature in Ref.~\cite{Lamberto2025}. For all the temperatures considered, the lower polariton is overdamped in a finite interval of couplings around $g_c(T)$. Increasing the temperature shifts the transition to stronger couplings and, beyond $g_c(T)$, accelerates the decoupling of light and matter. We then compute the reflection and thermal emission spectra in both phases. The emission becomes brighter close to $g_c(T)$, and when the dissipation rates of the two channels are suitably balanced, the system shows reflectionless modes~\cite{Sun2014,Sweeney2020,Bonizzoni2026}, at which it behaves simultaneously as a perfect absorber and a perfect emitter. Since the description remains valid across the transition, it may also be useful for criticality-enhanced sensing~\cite{Chu2021,Ilias2022,Hotter2024}, where the output fields near the critical point must be described accurately.

The paper is organized as follows. In \secref{sec:finiteT} we study the Dicke model at finite temperature, deriving the thermal mean-field state and the critical line (\secref{sec:thermaleq}) and the Hamiltonian of the fluctuations about the thermal mean field (\secref{sec:noneq_excitatoins}). In \secref{sec:langevin} we couple the system to the reservoirs and derive the quantum Langevin equations in the normal (\secref{sec:langevin-NP}) and superradiant (\secref{sec:superradiant}) phases. \mbox{Section~\ref{sec:io}} presents the input-output relations and the scattering amplitudes that determine the reflection and thermal emission spectra. These spectra are discussed for ohmic baths in \secref{sec:spectra}. Finally, \secref{sec:conclusions} is devoted to our conclusions.

\section{Dicke model at finite temperature}
\label{sec:finiteT}

\subsection{Thermal mean-field state and critical curve}
\label{sec:thermaleq}

\begin{figure}[t]
\centering
\includegraphics[width=\columnwidth]{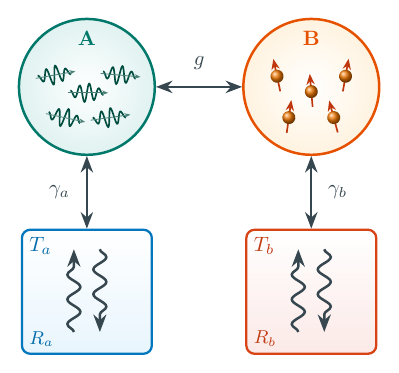}
\caption{\textbf{Sketch of the system.}  A single cavity mode (subsystem A, resonance frequency $\omega_a$) interacts with an ensemble of $N$ identical two-level systems (subsystem B, transition frequency $\omega_b$) via a collective coupling strength $g$. Each component is coupled to its own independent bosonic reservoir, with the photonic bath kept at zero temperature ($T_a = 0$) and the matter bath thermalized at a finite temperature ($T_b = T$). These system-bath interfaces also serve as input-output ports through which the  system can be excited and probed.}
\label{fig:system_sketch}
\end{figure}

In this section, we investigate the finite-temperature properties of the Dicke model, focusing first on the static thermodynamic phase diagram before explicitly incorporating the microscopic bath degrees of freedom and the resulting dissipative dynamics in Sec.~\ref{sec:langevin}.
As schematically illustrated in Fig.~\ref{fig:system_sketch}, we consider a configuration where the photonic reservoir is kept at zero temperature ($T_a = 0$), while the atomic ensemble is thermalized with its own reservoir at a finite temperature $T_b = T$. Despite the resulting non-equilibrium setting, we assume that the cavity does not alter the thermal state of the quantum emitters. This assumption is well justified: for a large ensemble, most of the spin degrees of freedom are decoupled from the cavity field (see, \textit{e.g.}, Refs.~\cite{Pino_2015, DallaTorre2016, Zappa2025cavity}), and a sufficiently strong interaction with the matter reservoir enforces thermalization at temperature $T$.
The coupled light-matter system consists of a single bosonic cavity mode of resonance frequency $\omega_a$ interacting with a collection of $N$ identical two-level systems  with transition frequency $\omega_b$. Introducing the collective spin operators $J_{x,y,z} = \tfrac{1}{2}\sum_{k=1}^N \sigma^{(k)}_{x,y,z}$, the system Hamiltonian is written as
\begin{equation}
H_{\rm sys} = \hbar\omega_a a^\dagger a + \hbar\omega_b J_z + \frac{2\hbar g}{\sqrt N}(a+a^\dagger)J_x \, ,
\label{eq:Hdicke}
\end{equation}
where $a$ and $a^\dagger$ denote the photon annihilation and creation operators, and $g$ is the collective light-matter coupling strength. Rescaling the bare coupling by an appropriate power of $N$ is standard practice for models with all-to-all interactions, such as the Lipkin-Meshkov-Glick model, and is required to keep the interaction energy extensive as $N\to\infty$; the factor $N^{-1/2}$ in \eqref{eq:Hdicke} serves precisely this purpose.

In the thermodynamic limit the cavity field is expected to develop a macroscopic coherent component in the superradiant phase, of the same physical origin as the macroscopic occupation of the zero-momentum mode in Bose-Einstein condensation. We formalize this by writing
\begin{equation}
a \longrightarrow \sqrt N \alpha \, , \qquad \alpha\in\mathbb R \, ,
\label{eq:alphadef}
\end{equation}
where $\alpha$ is an intensive, $\mathcal O(1)$ quantity as $N\to\infty$. We select a symmetry-broken equilibrium branch whenever the free energy develops two degenerate minima. The thermodynamic limit is taken at fixed coupling, frequencies, and temperature, and the critical behavior is subsequently obtained by approaching the transition on a stable branch.

Substituting \eqref{eq:alphadef} into \eqref{eq:Hdicke} gives the mean-field Hamiltonian
\begin{equation}
H_{\rm MF}(\alpha) = N\hbar\omega_a\alpha^2 + \hbar\omega_b J_z + 4\hbar g\alpha J_x \, ,
\label{eq:HMF}
\end{equation}
which still acts on the full atomic Hilbert space: only the cavity field has been replaced by its mean value, and no approximation has been made on the atoms; self-consistency on $\alpha$ is imposed only afterward. Since $J_i = \tfrac12\sum_k \sigma_i^{(k)}$, the spin part of \eqref{eq:HMF} decouples exactly into $N$ independent and identical single-atom Hamiltonians
\begin{equation}
h(\alpha) = \tfrac12 B_{\rm eff}(\alpha)\cdot\bm\sigma \, , \qquad B_{\rm eff}(\alpha) = (4\hbar g\alpha,\,0,\,\hbar\omega_b) \, .
\label{eq:Beff}
\end{equation}
Each atom therefore behaves as a spin-$\tfrac12$ particle in the effective field $B_{\rm eff}(\alpha)$: its longitudinal component is fixed by the bare atomic splitting $\omega_b$, while its transverse component is generated by the coherent cavity amplitude and is present only when $\alpha\neq0$. The field magnitude,
\begin{equation}
\epsilon(\alpha) \equiv |B_{\rm eff}(\alpha)| = \hbar\sqrt{\omega_b^2+(4g\alpha)^2} \equiv \hbar\Omega(\alpha) \, ,
\label{eq:epsilon}
\end{equation}
gives the eigenvalues $E_\pm=\pm\epsilon(\alpha)/2$ of $h(\alpha)$. For the canonical Gibbs state $\rho(\alpha)=e^{-\beta h(\alpha)}/z(\alpha)$, with $\beta\equiv1/k_BT$ and single-atom partition function $z(\alpha)=2\cosh[\beta\epsilon(\alpha)/2]$. Moreover the thermal average of the Pauli vector is $\langle\bm\sigma\rangle=-\hat n\tanh[\beta\epsilon(\alpha)/2]$, where $\hat n=B_{\rm eff}(\alpha)/\epsilon(\alpha)$ is the unit vector along the effective field; the derivation is given in full in \appref{app:spinavg}. Multiplying by $N/2$ gives the collective averages
\begin{equation}
\begin{aligned}
\langle J_z\rangle(\alpha) &= -\frac N2\frac{\omega_b}{\Omega(\alpha)}\tanh\!\left(\frac{\beta\epsilon(\alpha)}2\right) \, , \\
\langle J_x\rangle(\alpha) &= -\frac N2\frac{4g\alpha}{\Omega(\alpha)}\tanh\!\left(\frac{\beta\epsilon(\alpha)}2\right) \, ,
\end{aligned}
\label{eq:Jz}
\end{equation}
with $\langle J_y\rangle=0$. In the normal phase ($\alpha=0$) the atoms see only the bare longitudinal field, so $\langle J_x\rangle=0$ identically and \eqref{eq:Jz} reduces at $T=0$ to the fully polarized state $\langle J_z\rangle=-N/2$. As soon as $\alpha\neq0$, the transverse field tilts the spin toward $x$, while raising the temperature suppresses both components through the common thermal factor $\tanh[\beta\epsilon(\alpha)/2]$, which interpolates between full polarization at $T=0$ and complete loss of coherence as $T\to\infty$.

The equilibrium value of $\alpha$ is obtained by minimizing the mean-field free energy, $F_{\rm MF}(\alpha)=-\beta^{-1}\ln Z_{\rm MF}(\alpha)$, with $Z_{\rm MF}(\alpha)={\rm Tr}\,e^{-\beta H_{\rm MF}(\alpha)}$. For this purpose we use the exact derivative identity
\begin{equation}
\frac{\partial F_{\rm MF}}{\partial\alpha} = \left\langle\frac{\partial H_{\rm MF}}{\partial\alpha}\right\rangle_{\rho_{\rm MF}} \, ,
\label{eq:HF}
\end{equation}
which is the thermal counterpart of the Hellmann--Feynman theorem and is derived in \appref{app:stability}. Differentiating \eqref{eq:HMF} gives $\partial_\alpha H_{\rm MF}=2N\hbar\omega_a\alpha+4\hbar gJ_x$, so that $\partial_\alpha F_{\rm MF}=0$ is equivalent to
\begin{equation}
\omega_a\alpha = -\frac{2g}N\langle J_x\rangle(\alpha) \, .
\label{eq:selfcons0}
\end{equation}
This is the essence of mean-field self-consistency: the amplitude entering $H_{\rm MF}$ must be reproduced by the thermal state that $H_{\rm MF}$ itself generates. Substituting \eqref{eq:Jz} into \eqref{eq:selfcons0} gives the transcendental self-consistency equation
\begin{equation}
\alpha\left[\omega_a - \frac{4g^2}{\Omega(\alpha)}\tanh\!\left(\frac{\beta\epsilon(\alpha)}2\right)\right] = 0 \, ,
\label{eq:selfcons}
\end{equation}
whose numerical solutions determine the equilibrium order parameter $\alpha(g,\beta)$ at every coupling and temperature, exhibiting the mean-field bifurcation characteristic of a continuous transition. The two equilibrium phases correspond to the vanishing or nonzero root of \eqref{eq:selfcons}, which we treat in turn below.

In the normal phase the coherent cavity amplitude vanishes, $\alpha=0$, so that $\Omega(0)=\omega_b$ and \eqref{eq:Jz} reduces to
\begin{equation}
\langle J_z\rangle = -\frac N2\tanh\!\left(\frac{\beta\hbar\omega_b}2\right) \, , \qquad \langle J_x\rangle = 0 \, .
\label{eq:Jz-NP}
\end{equation}
Because $\langle J_x\rangle=0$ identically at $\alpha=0$, \eqref{eq:selfcons0} is satisfied trivially at every coupling and temperature, and $\alpha=0$ is always a stationary point of the free energy.

To extract the critical line analytically, we take the nonzero branch of \eqref{eq:selfcons}, divide by $\alpha$, and send $\alpha\to0^+$, so that $\Omega(\alpha)\to\omega_b$. This gives the closed-form critical curve
\begin{equation}
g_c(T)^2 = \frac{\omega_a\omega_b}4\coth\!\left(\frac{\beta\hbar\omega_b}2\right) \, .
\label{eq:gcT}
\end{equation}
For $g < g_c(T)$, the system is in the normal phase. To verify that the stationary point at $\alpha=0$ is a true free-energy minimum, we expand $F_{\mathrm{MF}}(\alpha)$ to second order about the origin (see \appref{app:stability}). The curvature is strictly positive for $g < g_c(T)$ but changes sign exactly at $g = g_c(T)$, marking the appearance of two new symmetric minima at $\pm\alpha(g,T)$.

At $T=0$, $\coth(\beta\hbar\omega_b/2)\to1$ and \eqref{eq:gcT} recovers the zero-temperature critical coupling $g_c(0)=\sqrt{\omega_a\omega_b}/2$ of the closed Dicke model. In the opposite limit, $\beta\hbar\omega_b\ll1$, the asymptotic expansion $\coth x\simeq1/x$ gives $g_c(T)\simeq\sqrt{\omega_ak_BT/(2\hbar)}$, which diverges as $T\to\infty$: strong thermal fluctuations suppress the net atomic polarization, and an arbitrarily large collective coupling is required to overcome this randomization and establish a macroscopic cavity field.

\begin{figure}[t]
\centering
\includegraphics[width=\columnwidth]{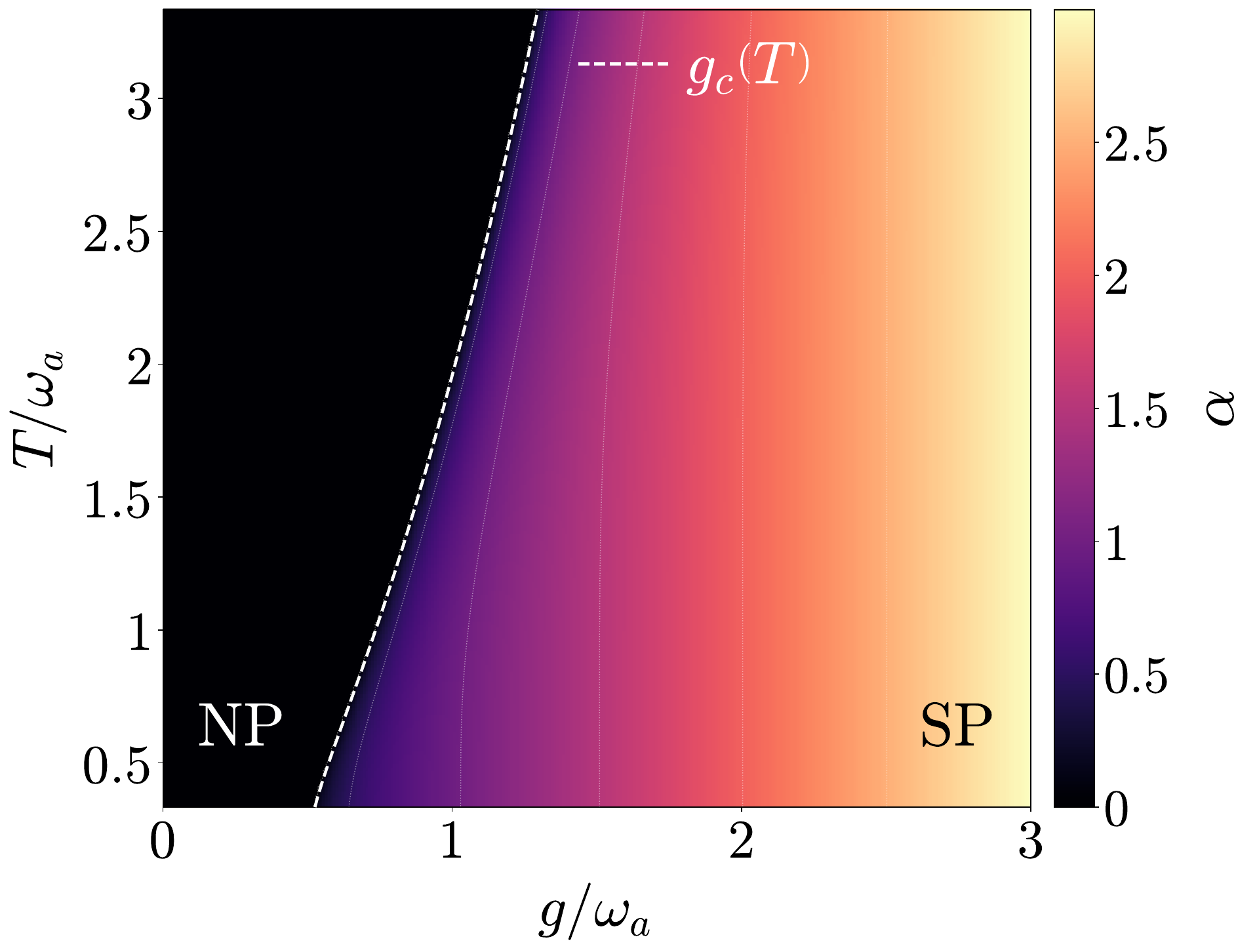}
\caption{Equilibrium order parameter $\alpha(g,T)$, obtained by numerically solving the self-consistency equation, \eqref{eq:selfcons}, choosing the branch $\alpha\ge0$, as a function of the normalized coupling strength and of $T$. The dashed line marks the critical line $g_c(T)$ of \eqref{eq:gcT}. The order parameter vanishes throughout the normal phase (NP) and grows continuously into the superradiant phase (SP) beyond the transition.}
\label{fig:alphacontour}
\end{figure}

Above the critical coupling the equilibrium field acquires a nonzero amplitude and produces a transverse component of the atomic effective field, with $\alpha$ determined by the nonzero-$\alpha$ branch of \eqref{eq:selfcons},
\begin{equation}
\omega_a = \frac{4g^2}{\Omega(\alpha)}\tanh\!\left(\frac{\beta\hbar\Omega(\alpha)}2\right) \, .
\label{eq:selfcons-SP}
\end{equation}
At fixed coupling and temperature, we solve \eqref{eq:selfcons-SP} numerically to determine the equilibrium amplitude $\alpha$. This amplitude fixes the direction and magnitude of the effective atomic field, from which the thermal spin polarization follows via \eqref{eq:Jz}. {\renewcommand{\figref}[1]{\mbox{Figure~\ref{#1}}}\figref{fig:alphacontour}} shows the resulting order parameter across the coupling-temperature plane, choosing the branch $\alpha\ge0$: it vanishes throughout the normal-phase region and grows continuously beyond the critical line, with the numerically located boundary coinciding, at every temperature shown, with the closed-form critical line of \eqref{eq:gcT}.

\subsection{Fluctuation Hamiltonian}
\label{sec:noneq_excitatoins}

Having determined the equilibrium amplitude and polarization, we express the operator Hamiltonian in terms of deviations from these mean values. Using $J_x=(J_++J_-)/2$, \eqref{eq:Hdicke} reads
\begin{equation}
H_{\rm sys} = \hbar\omega_a a^\dagger a + \hbar\omega_b J_z + \frac{\hbar g}{\sqrt N}(a+a^\dagger)(J_++J_-) \, ,
\label{eq:Hsys2}
\end{equation}
and the cavity operators are displaced as $a=\sqrt N\alpha+\delta a$, with $\langle\delta a\rangle=0$. For the atomic ensemble the effective field lies in the $xz$ plane, so a rotation about $y$,
\begin{equation}
\begin{pmatrix}J_x\\J_z\end{pmatrix} = \begin{pmatrix}\cos\theta&\sin\theta\\-\sin\theta&\cos\theta\end{pmatrix}\begin{pmatrix}J_x'\\J_z'\end{pmatrix} \, , \quad J_y=J_y' \, ,
\label{eq:rotation}
\end{equation}
brings the new $z'$ axis into alignment with it. Substituting \eqref{eq:rotation} into the mean-field spin energy, $\omega_bJ_z+4g\alpha J_x=(\omega_b\cos\theta+4g\alpha\sin\theta)J_z'+(4g\alpha\cos\theta-\omega_b\sin\theta)J_x'$, the coefficient of $J_x'$ vanishes for
\begin{equation}
\tan\theta = \frac{4g\alpha}{\omega_b} \, , \qquad \overline\omega_b \equiv \Omega(\alpha) = \sqrt{\omega_b^2+(4g\alpha)^2} \, ,
\label{eq:theta}
\end{equation}
leaving the spin energy $\hbar\overline\omega_bJ_z'$. The Gibbs state is therefore diagonal in the rotated basis, and the single-atom calculation leading to \eqref{eq:Jz}, now applied with the field along $z'$, gives $\langle J_x'\rangle=\langle J_y'\rangle=0$ and $\langle J_z'\rangle=-\tfrac N2\tanh(\beta\hbar\overline\omega_b/2)$.

We complete the expansion by separating the mean population along $z'$ from its fluctuation, $J_z'=\langle J_z'\rangle+\sqrt N L$ ~\cite{Goderis1989,Goderis1990,Boneberg2022}. 
Notice that $\langle J^\prime_x\rangle=\langle J^\prime_y\rangle=0 $. The original spin components then read
\begin{align}
J_x &= \langle J_x\rangle + \cos\theta\, J_x' + \sqrt N\sin\theta\, L \, ,\\
J_z &= \langle J_z\rangle - \sin\theta\, J_x' + \sqrt N\cos\theta\, L \, ,
\end{align}
with $\langle J_x\rangle=\sin\theta\langle J_z'\rangle$ and $\langle J_z\rangle=\cos\theta\langle J_z'\rangle$. In the mean-field thermal state, the centered collective spin fluctuations scale as $J_x',J_y'=\mathcal O(\sqrt N)$, while $L=\mathcal O(1)$ and $\delta a=\mathcal O(1)$ (\appref{app:expansion}). Substituting these decompositions, together with $a=\sqrt N\alpha+\delta a$, into \eqref{eq:Hsys2}, and separating the additive mean-field energy $E_{\rm MF}=N\hbar\omega_a\alpha^2+\hbar\overline\omega_b\langle J_z'\rangle$ from the terms linear in $\delta a+\delta a^\dagger$ and the remaining operator terms, gives
\begin{align}
H_{\rm sys} = {}&E_{\rm MF} + \hbar\omega_a\delta a^\dagger\delta a + \hbar\overline\omega_b\sqrt N L \notag\\
&+ \hbar\sqrt N\left(\omega_a\alpha+\frac{2g\sin\theta}N\langle J_z'\rangle\right)(\delta a+\delta a^\dagger) \notag\\
&+ \frac{\hbar g\cos\theta}{\sqrt N}(\delta a+\delta a^\dagger)(J_+'+J_-') \notag\\
&+ 2\hbar g\sin\theta(\delta a+\delta a^\dagger)L \, .
\end{align}
Since $\sin\theta\langle J_z'\rangle=\langle J_x\rangle$, the coefficient of $\delta a+\delta a^\dagger$ is exactly $\omega_a\alpha+(2g/N)\langle J_x\rangle$, which vanishes by the self-consistency condition, \eqref{eq:selfcons0}: the cancellation of every term of order $\sqrt N$ is thus a direct consequence of expanding about the equilibrium mean field, and not a further assumption. Dropping the additive constant $E_{\rm MF}$ leaves the fluctuation Hamiltonian
\begin{equation}
\begin{aligned}
H_{\rm sys} &= \hbar\omega_a\delta a^\dagger\delta a + \hbar\overline\omega_b\sqrt N L \\
&+ \frac{\hbar g\cos\theta}{\sqrt N}(\delta a+\delta a^\dagger)(J_+'+J_-') \\
&+ 2\hbar g\sin\theta(\delta a+\delta a^\dagger)L \, ,
\end{aligned}
\label{eq:HSPfluct}
\end{equation}
with
\begin{equation}
\cos\theta=\frac{\omega_b}{\overline\omega_b} \, , \qquad \sin\theta=\frac{4g\alpha}{\overline\omega_b} \, .
\label{eq:thetacossin}
\end{equation}
Both interaction terms in \eqref{eq:HSPfluct} are $\mathcal O(1)$ and must be retained: the transverse term couples the field to spin fluctuations perpendicular to the mean polarization, while the longitudinal term arises because population fluctuations along the tilted axis retain a component along the original coupling direction $J_x$. For $\alpha=0$ the displacement and rotation vanish, $\overline\omega_b=\omega_b$, and \eqref{eq:HSPfluct} reduces to the normal-phase Hamiltonian up to the subtracted mean-field energy.

\section{Open Dicke model with dissipation: quantum Langevin equations}
\label{sec:langevin}

We now extend the thermal mean-field description of \secref{sec:finiteT} by coupling the system to an external environment. We develop a unified theoretical framework, valid for both the normal and superradiant phases, that places no restriction on the system-bath coupling strength or the frequency dependence of the dissipative kernel. This non-perturbative approach is  necessary in the vicinity of the critical point, where standard weak-coupling approximations fail. 

We model the environment by coupling the cavity and the collective matter coordinates to two independent bosonic reservoirs, each consisting of an infinite collection of harmonic oscillators. The total Hamiltonian, expressed in terms of the original, unshifted operators, reads
\begin{equation}
H_{\rm tot} = H_{\rm sys} + \sum_{j=a,b}\sum_n \left[ \frac{p_{jn}^{2}}{2m_{jn}} + \frac{k_{jn}}{2}(q_{jn}-X_j)^2 \right] \, ,
\label{eq:Htot_unshifted}
\end{equation}
where the system coordinates are
\begin{equation}
X_a = \sqrt{\frac{\hbar}{2\omega_a}}(a+a^\dagger) \, , \qquad X_b = \sqrt{\frac{\hbar}{2\omega_b}} \frac{J_++J_-}{\sqrt N} \, ,
\label{eq:original_system_coordinates}
\end{equation}
and the bare reservoir frequencies satisfy $\omega_{jn}^{2}=k_{jn}/m_{jn}$. This specific interaction potential features a metastable energy minimum; a purely bilinear alternative, $q_{jn}X_j$, lacks this feature and artificially shifts the location of the transition~\cite{Lamberto2025,musolino2026reservoir}.
An equivalent, Fano-Hopfield-type diagonalization~\cite{SavastaGirlanda1996,Cortese2022} of the full system-bath Hamiltonian is possible but is not pursued here, since the Heisenberg-Langevin route taken below gives direct access to the input-output relations of \secref{sec:io}.

In the superradiant phase, the emergence of macroscopic mean fields displaces the reservoirs. Imposing stationarity with respect to the bath coordinates yields $\langle q_{jn}\rangle=\langle X_j\rangle$ and $\langle p_{jn}\rangle=0$. At these stationary values, the linear reservoir potential vanishes, leaving the equilibrium conditions derived in \secref{sec:finiteT}. Because the interaction energy vanishes identically at this stationary point, eliminating the reservoirs first and then extremizing over the system coordinates is equivalent to extremizing the full Hamiltonian $H_{\rm tot}$ over system and bath variables at once: the reservoirs exert no back-action on the stationarity condition for $X_j$. This equivalence relies on the shifted-oscillator form of \eqref{eq:Htot_unshifted}. A bilinear coupling $q_{jn}X_j$ would instead leave a nonvanishing, $X_j$-dependent term after the bath is eliminated; this term feeds back into the same condition and is responsible for the spurious shift of the transition noted above. Expanding each coordinate around its equilibrium value, $q_{jn}-X_j = \delta q_{jn}-\delta X_j$, we find $\delta X_a = X_{{\rm s},a}$ for the cavity. For the matter coordinate, the spin rotation and the relation $\overline\omega_b=\omega_b/\cos\theta$ yield
\begin{equation}
\delta X_b = \sqrt{\frac{\hbar}{2\omega_b}} \left[ \cos\theta\,\frac{J_+'+J_-'}{\sqrt N} + 2\sin\theta\,L \right] = \sqrt{\cos\theta}\,X_{{\rm s},b} \, ,
\label{eq:physical_matter_coordinate}
\end{equation}
where $X_{{\rm s},a}$ and $X_{{\rm s},b}$ denote the shifted system operators (specified explicitly in \eqref{eq:Xs}).

To restore the canonical quadratic structure of the matter reservoir, we apply the coordinate transformation
\begin{equation}
\begin{aligned}
q_{{\rm s,a}n}&=\delta q_{{\rm a}n}, & p_{{\rm s,a}n}&=\delta p_{{\rm a}n}, \\
q_{{\rm s,b}n}&=\frac{\delta q_{{\rm b}n}}{\sqrt{\cos\theta}}, & p_{{\rm s,b}n}&=\sqrt{\cos\theta}\,\delta p_{{\rm b}n} \, .
\end{aligned}
\label{eq:bath_canonical_transformation}
\end{equation}
This transformation leaves the cavity reservoir unchanged but rescales the matter-bath parameters to $M_{bn}=m_{bn}\cos\theta$ and $\tilde k_{{\rm s},bn}=k_{bn}\cos\theta$. The bare oscillator frequencies remain invariant since $\tilde k_{{\rm s},bn}/M_{bn}=k_{bn}/m_{bn}$.  
This rescaling demonstrates that the effective coupling between the rotated matter coordinate and its reservoir is multiplied by the geometric factor $\cos\theta$ (which trivially reduces to unity in the normal phase where $\theta=0$). The shifted interaction Hamiltonian thus becomes
\begin{equation}
H_{\rm int} = \sum_{j=a,b}\sum_n \left[ \frac{p_{{\rm s},jn}^{2}}{2M_{jn}} + \frac{\tilde k_{{\rm s},jn}}2(q_{{\rm s},jn}-X_{s,j})^2 \right] \, .
\label{eq:Hint}
\end{equation}

Defining the reference memory kernels $f_j(t) = \sum_n k_{jn}\cos(\omega_{jn}t)$, the effective kernel for the transformed matter reservoir is correspondingly scaled by $\cos\theta$. Integrating the Heisenberg equations of motion for the bath oscillators and substituting the results back into the system dynamics eliminates the reservoir degrees of freedom. Sending the preparation time to the remote past ($t_0\to-\infty$), we obtain the exact quantum Langevin equation (\appref{app:E}):
\begin{equation}
\begin{aligned}
\dot Y &= \frac i\hbar[H_{\rm sys},Y]\\
&\quad - \sum_{j=a,b}\frac{i}{2\hbar} \Bigl\{ [X_{{\rm s},j},Y], \, \xi_j(t)\\
&\qquad - \int_{-\infty}^t dt'\, f_j(t-t') \dot X_{{\rm s},j}(t') \Bigr\}_+ \, ,
\end{aligned}
\label{eq:langevin-general}
\end{equation}
where $\xi_j(t)$ represents the free reservoir noise, and the kernel for the matter reservoir is understood as $f_b(t)\to\cos\theta\,f_b(t)$, in accordance with the rescaling above. We emphasize that no Born, Markov, or rotating-wave approximations are invoked in this reservoir elimination. 

Evaluating \eqref{eq:langevin-general} for the cavity field and the lowering operator $J_-'$ yields
\begin{align}
\delta\dot a ={}& -i\omega_a\delta a -i\bar g(J_+'+J_-') -2ig\sin\theta\,L \notag\\
&- \frac{i}{2\omega_a} \int_{-\infty}^t\!dt'\, f_a(t-t') [\delta\dot a^\dagger(t')+\delta\dot a(t')] \notag\\
&+ \frac{i}{\sqrt{2\hbar\omega_a}}\xi_a(t) \, , \label{eq:dota} \\
\dot J_-' ={}& -i\overline\omega_b J_-' + 2i\bar g\langle J_z'\rangle (\delta a+\delta a^\dagger) \notag\\
&+ \frac{i\langle J_z'\rangle\cos\theta}{N\overline\omega_b} \int_{-\infty}^t\!dt'\, f_b(t-t') [\dot J_+'(t')+\dot J_-'(t')] \notag\\
&- i\sqrt{\frac2{N\hbar\overline\omega_b}} \langle J_z'\rangle\,\xi_b(t) \, , \label{eq:dotJminus}
\end{align}
where $\bar g\equiv g\cos\theta/\sqrt N$. The conjugate equations for $\delta a^\dagger$ and $J_+'$ follow by Hermitian conjugation. The longitudinal fluctuation $L$ is dynamically decoupled in the thermodynamic limit: its coherent and dissipative couplings scale as $\mathcal O(N^{-1/2})$, yielding $\dot L \to 0$. In the Fourier  domain we thus set $\tilde L(\omega)=0$ for $\omega\neq0$. This restriction applies exclusively to the dynamical response and leaves the static thermal population, $J_z'=\langle J_z'\rangle+\sqrt N L$, physically intact.

Fourier transforming the remaining equations, with the convention $\tilde Y(\omega)=\int dt\, e^{i\omega t}Y(t)$, casts the dynamics of the fluctuation vector $v=(\delta\tilde a,\delta\tilde a^\dagger,\tilde J_-',\tilde J_+')^T$ into the compact form
\begin{equation}
-i\omega\,v(\omega) = -i\left(A-\frac i2\Gamma(\omega)\right)v(\omega) + \tilde F_{\rm in}(\omega) \, .
\label{eq:langevinFourier}
\end{equation}
Here, the Langevin-force vector is
\begin{equation}
\tilde F_{\rm in}(\omega) = 
\begin{pmatrix}
\displaystyle \frac{i}{\sqrt{2\hbar\omega_a}}\, \tilde\xi_a(\omega) \\[6pt]
\displaystyle -\frac{i}{\sqrt{2\hbar\omega_a}}\, \tilde\xi_a(\omega) \\[6pt]
\displaystyle -i\sqrt{\frac{2}{N\hbar\overline\omega_b}}\, \langle J_z'\rangle\,\tilde\xi_b(\omega) \\[6pt]
\displaystyle i\sqrt{\frac{2}{N\hbar\overline\omega_b}}\, \langle J_z'\rangle\,\tilde\xi_b(\omega)
\end{pmatrix} \, ,
\label{eq:Fin}
\end{equation}
the Hopfield-Bogoliubov matrix reads
\begin{equation}
\begingroup
\setlength{\arraycolsep}{6pt}
A = \begin{pmatrix}
\omega_a & 0 & \bar g & \bar g \\
0 & -\omega_a & -\bar g & -\bar g \\
-2\bar g\langle J_z'\rangle & -2\bar g\langle J_z'\rangle & \overline\omega_b & 0 \\
2\bar g\langle J_z'\rangle & 2\bar g\langle J_z'\rangle & 0 & -\overline\omega_b
\end{pmatrix} \, ,
\endgroup
\label{eq:Amatrix}
\end{equation}
and the frequency-dependent decay matrix is
\begin{equation}
\begingroup
\setlength{\arraycolsep}{2pt}
\Gamma(\omega) = \begin{pmatrix}
\gamma_a(\omega) & -\gamma_a(\omega) & 0 & 0 \\
-\gamma_a(\omega) & \gamma_a(\omega) & 0 & 0 \\
0 & 0 & \tilde\gamma_b(\omega) & -\tilde\gamma_b(\omega) \\
0 & 0 & -\tilde\gamma_b(\omega) & \tilde\gamma_b(\omega)
\end{pmatrix} \, .
\endgroup
\label{eq:Gammaomega}
\end{equation}
The quantity $\gamma_j(\omega)$ is defined as the Fourier transform of the causal kernel $\theta(t)f_j(t)$. Because $f_j(t)$ is real, this transform satisfies $\gamma_j^*(\omega)=\gamma_j(-\omega)$; we treat it as real-valued throughout the following analysis. The matter block is governed by the decay rate $\tilde\gamma_b(\omega) \equiv \gamma_{b,\rm n}(\omega)\cos\theta$, where the polarization-dressed matter decay rate
\begin{equation}
\gamma_{b,\rm n}(\omega) \equiv -\frac{2\langle J_z'\rangle}{N}\gamma_b(\omega)\,,
\label{eq:gamma_beff}
\end{equation}
incorporates the thermal polarization $\langle J_z'\rangle$ on top of the bare rate $\gamma_b(\omega)$, and reduces to $\gamma_{b,\rm n}(\omega)=\gamma_b(\omega)$ at $T=0$. {\renewcommand{\eqref}[1]{\mbox{(\ref{#1})}}Equations~\eqref{eq:Amatrix} and \eqref{eq:Gammaomega}} already hold in either phase, through their dependence on $\theta$; we now specialize them in turn.

\subsection{Normal phase}
\label{sec:langevin-NP}
In the normal phase $\alpha=0$, so that $\theta=0$, $\overline\omega_b=\omega_b$, and $\langle J_z'\rangle=\langle J_z\rangle$. The coherent-dynamics and decay matrices of {\renewcommand{\eqref}[1]{\mbox{(\ref{#1})}}Eqs.~\eqref{eq:Amatrix} and \eqref{eq:Gammaomega}} reduce to
\begin{equation}
\begingroup
\setlength{\arraycolsep}{6pt}
\begin{aligned}
A_{\rm NP} &= \begin{pmatrix}
\omega_a & 0 & g_0 & g_0 \\
0 & -\omega_a & -g_0 & -g_0 \\
-2g_0\langle J_z\rangle & -2g_0\langle J_z\rangle & \omega_b & 0 \\
2g_0\langle J_z\rangle & 2g_0\langle J_z\rangle & 0 & -\omega_b
\end{pmatrix} \, , \\[12pt]
\Gamma_{\rm NP}(\omega) &= \begin{pmatrix}
\gamma_a(\omega) & -\gamma_a(\omega) & 0 & 0\\
-\gamma_a(\omega) & \gamma_a(\omega) & 0 & 0\\
0&0& \gamma_{b,\rm n}(\omega) & -\gamma_{b,\rm n}(\omega)\\
0&0& -\gamma_{b,\rm n}(\omega) & \gamma_{b,\rm n}(\omega)
\end{pmatrix} \, ,
\end{aligned}
\endgroup
\label{eq:ANPGammaNP}
\end{equation}
with $g_0\equiv g/\sqrt N$ and $\langle J_z\rangle$ fixed by its finite-temperature value (see \eqref{eq:Jz-NP}). 
Defining $M(\omega;A,\Gamma)\equiv A-i\Gamma(\omega)/2-\omega I$, the complex eigenfrequencies of the system follow from
\begin{equation}
\zeta(\omega) \equiv \det M(\omega;A,\Gamma) = 0 \, ,
\label{eq:zeta}
\end{equation}
evaluated with $A_{\rm NP}$ and $\Gamma_{\rm NP}(\omega)$, before any assumption on the frequency dependence of $\Gamma_{\rm NP}(\omega)$ is made. As $\omega\to0$, the only surviving contribution to $\zeta_{\rm NP}(\omega)$ is its constant term, which vanishes precisely at $g=g_c(T)$ for any dissipation rate satisfying $\omega\gamma_j(\omega)\to0$ as $\omega\to0$: the finite-temperature critical curve of \eqref{eq:gcT} (see white dashed curve in \figref{fig:alphacontour} ) is therefore left unaffected by the coupling to the reservoirs, generalizing to finite temperature the resilience of the equilibrium transition already established at $T=0$~\cite{Lamberto2025}.
\subsection{Superradiant phase}\label{sec:superradiant}

In the superradiant phase the equilibrium nonzero mean system coordinates also displace the reservoirs; writing $c_{jn}=c_{s,jn}\pm\sqrt N\,\sigma_{jn}$ (the sign following the convention of \secref{sec:noneq_excitatoins}) and imposing stationarity of the bath coordinates against the system means gives the induced bath occupations
\begin{equation}
\overline\sigma_{an} = \sqrt{\frac{k_{an}}{\omega_{an}\omega_a}}\,\alpha \, , \qquad
\overline\sigma_{bn} = -\frac1N\sqrt{\frac{k_{bn}}{\omega_b\omega_{bn}}}\sin\theta\,\langle J_z'\rangle \, .
\label{eq:sigmabar}
\end{equation}
Expressed in terms of the fluctuation operators, the shifted reservoir Hamiltonian retains the same structure as \eqref{eq:Hint}, with shifted system coordinates
\begin{equation}
\begin{aligned}
X_{s,a} &= \sqrt{\frac{\hbar}{2\omega_a}}(\delta a+\delta a^\dagger) \, , \\
X_{s,b} &= \sqrt{\frac{\hbar}{2\overline\omega_b}}\left[\frac{J_+'+J_-'}{\sqrt N}+2\tan\theta\,L\right] \, ,
\end{aligned}
\label{eq:Xs}
\end{equation}

The quantum Langevin equations retain the same structure as {\renewcommand{\eqref}[1]{\mbox{(\ref{#1})}}Eqs.~\eqref{eq:langevinFourier}--\eqref{eq:Gammaomega}}, now evaluated at $\theta\neq0$. We denote the resulting matrices $A_{\rm SP}\equiv A$ and $\Gamma_{\rm SP}(\omega)\equiv\Gamma(\omega)$. They are related to their normal-phase counterparts, \eqref{eq:ANPGammaNP}, by the parameter substitutions $\omega_b\to\overline\omega_b$ and $g\to g\cos\theta$, alongside the replacement of the normal thermal polarization $\langle J_z\rangle$ with the rotated one $\langle J_z'\rangle$. The matter block of the decay matrix acquires the extra rotation factor, mapping $\gamma_{b,\rm n}(\omega) \to \gamma_{b,\rm n}(\omega)\cos\theta \equiv \tilde\gamma_b(\omega)$. The  excitation energies again follow from $\zeta_{\rm SP}(\omega)\equiv\det M(\omega;A_{\rm SP},\Gamma_{\rm SP})=0$. The effective matter loss rate $\tilde\gamma_b(\omega)$ coincides with its normal-phase value $\gamma_{b,\rm n}(\omega)$ at the critical point (where $\theta\to0$) and is progressively suppressed deeper into the superradiant phase; this behavior, and its consequences for the emission spectra, are discussed quantitatively in \secref{sec:spectra}.

\begin{figure*}[t]
\centering
\includegraphics[width=\textwidth]{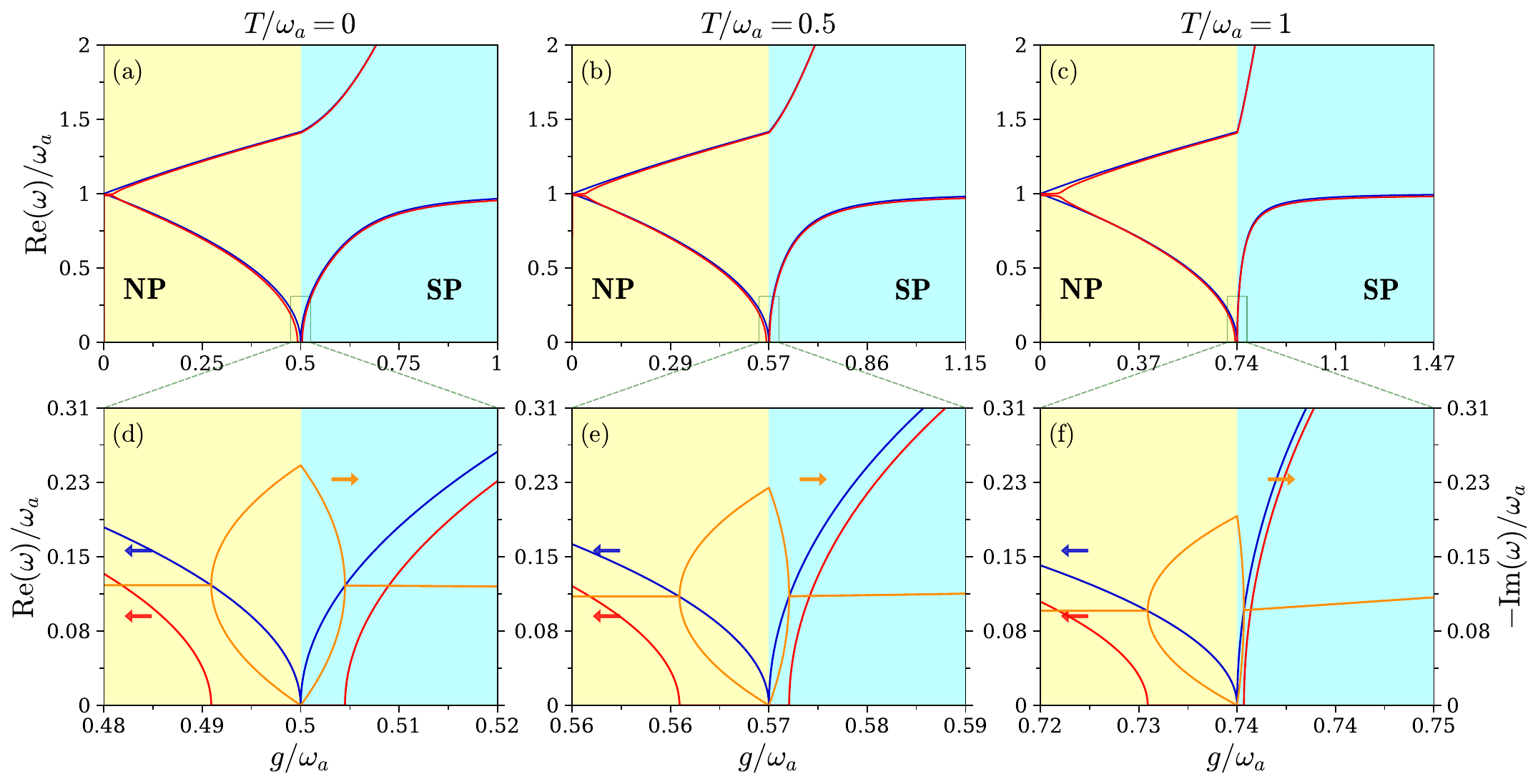}
\caption{\textbf{Collective spectrum.} \textbf{(a--c)} Upper and lower excitation energies as a function of the normalized coupling for the system without (blue) and with (red, real parts) coupling to the reservoirs, at $T/\omega_a=0,\,0.5,\,1$, respectively. \textbf{(d--f)} Insets zooming near the critical point onto the lower-polariton branch, for both the system with and without reservoir coupling.  The orange curve represents the negative
imaginary parts of the  lower  excitation energies. Parameters used: $\omega_a=\omega_b=1$, $\gamma_a=0.3$, $\gamma_b=0.2$.}
\label{fig:disp_rel}
\end{figure*}

{\renewcommand{\figref}[1]{\mbox{Figure~\ref{#1}}}
Figure~\ref{fig:disp_rel} displays the collective excitation spectrum determined from the complex roots of ~\eqref{eq:zeta}. In particular, we compare the real eigenfrequencies in the absence (blue curves) and presence (red curves) of reservoir coupling, alongside the corresponding damping rate of the lower polariton ($-\mathrm{Im}(\omega)/\omega_a$, orange curve).
Consistent with the zero-temperature analysis~\cite{Lamberto2025}, the low-frequency sector exhibits
the characteristic crossover from oscillatory to overdamped
dynamics as the coupling approaches the transition.
In the underdamped regime, the lower polariton has a complex
eigenfrequency with positive real part and negative imaginary part.
At the boundary of the overdamped region, the real part vanishes and
the eigenfrequency becomes purely imaginary, marking the analogue of
critical damping for a harmonic oscillator~\cite{dekker1981classical, musolino2026reservoir}. Beyond this boundary, it splits into
two purely imaginary eigenfrequencies describing fast and slow
relaxation. The resulting finite interval of coupling
strengths around $g_c(T)$, resolved in
Figs.~\ref{fig:disp_rel}(d--f), constitutes a dissipative
\emph{gap}: throughout this interval, the lower mode has no
oscillatory component. We identify the critical point of the phase transition with the simultaneous vanishing of both the real and imaginary parts of this excitation energy. Extending the normal-phase dispersion beyond this point would instead predict a positive imaginary part for the lower mode, in violation of causality, which forbids poles of a response function in the upper half of the complex-frequency plane. We confirm in \secref{sec:io} that the poles of the reflection coefficient $S_{11}(\omega)$ coincide with the roots of \eqref{eq:zeta}, validating this identification at any temperature.

Raising the temperature leaves this qualitative picture intact but reshapes it quantitatively in two distinct ways. First, the transition onset is delayed toward stronger couplings: as $g_c(T)$ increases according to \eqref{eq:gcT}, the normal-phase domain (yellow shaded area) progressively expands from {\mbox{Fig.~\ref{fig:disp_rel}}}(a) to Fig.~\ref{fig:disp_rel}(c). Second, the  emergence of superradiance becomes  steeper at higher temperatures. Rather than exhibiting the gradual zero-temperature growth away from $g_c$, the condensate amplitude $\alpha$ surges more rapidly over a narrow coupling interval as $T$ increases (see Fig.~\ref{fig:alphacontour}). This sharp growth of $\alpha$ has two simultaneous consequences through the self-consistent spin rotation of \secref{sec:noneq_excitatoins}: the effective light-matter coupling $\bar g=g\cos\theta/\sqrt N$ is rapidly suppressed as $\cos\theta\to0$, while the renormalized matter frequency $\overline\omega_b=\omega_b/\cos\theta$ grows correspondingly large. 
Consequently, once the critical threshold is crossed, the light and matter degrees of freedom decouple much more rapidly at elevated temperatures. The system  approaches the asymptotic regime of two independent harmonic oscillators: the lower branch tends to the bare cavity frequency as $\bar{g} \to 0$, whereas the upper branch tracks the diverging atomic scale $\overline{\omega}_b$. This thermally accelerated decoupling is directly reflected in the progressive steepening of the upper polariton branch above $g_c$ across {\renewcommand{\figref}[1]{\mbox{Figs.~\ref{#1}}}\figref{fig:disp_rel}}(a--c), as well as in the rapid flattening of the lower branch clearly visible in Fig.~\ref{fig:disp_rel}(c).

\begin{figure*}[htb]
    \centering
    \includegraphics[width=\textwidth]{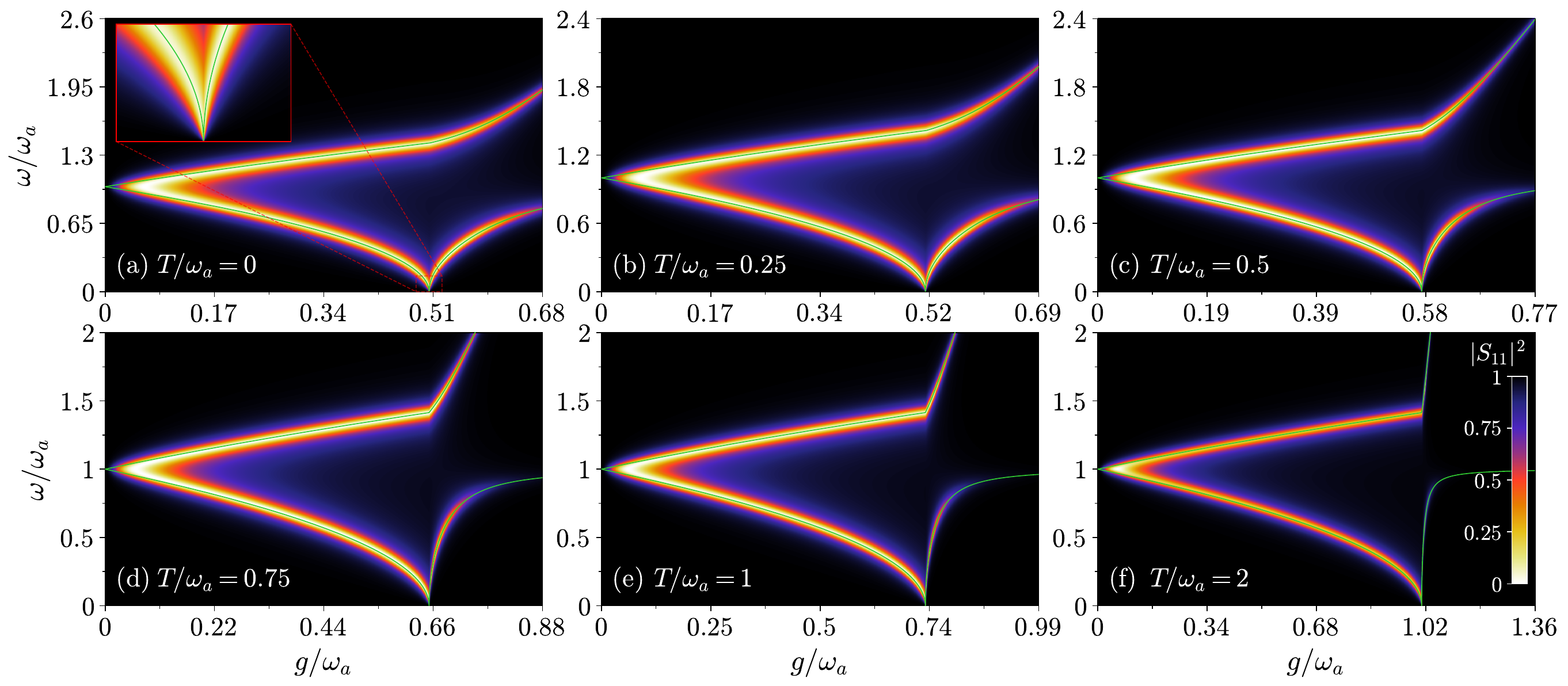}
    \caption{\textbf{Coherent cavity reflection spectra across the finite-temperature phase transition.}
    Color maps of the reflectivity $|S_{11}(\omega)|^2$ [\eqref{eq:S11}] as a function of the normalized coupling $g/\omega_a$ and probe frequency $\omega/\omega_a$ for an Ohmic bath at six temperatures: \textbf{(a)}~$T/\omega_a = 0$, \textbf{(b)}~$0.25$, \textbf{(c)}~$0.5$, \textbf{(d)}~$0.75$, \textbf{(e)}~$1$, and \textbf{(f)}~$2$. The green solid curves denote the polaritonic eigenfrequencies in the absence of reservoir coupling. The inset in panel~\textbf{(a)} zooms in on the zero-temperature critical point, $g_c(0)/\omega_a = 0.5$, showing the dissipative frequency shift vanishing as $\omega \to 0$. Parameters used: $\omega_a = \omega_b = 1$, $\gamma_a = 0.1$, $\gamma_b = 0.15$.}
    \label{fig:refl_spectra}
\end{figure*}
\section{Input--output fields: reflection and thermal emission spectra}
\label{sec:io}
\begin{figure*}[htb]
    \centering
    \includegraphics[width=\textwidth]{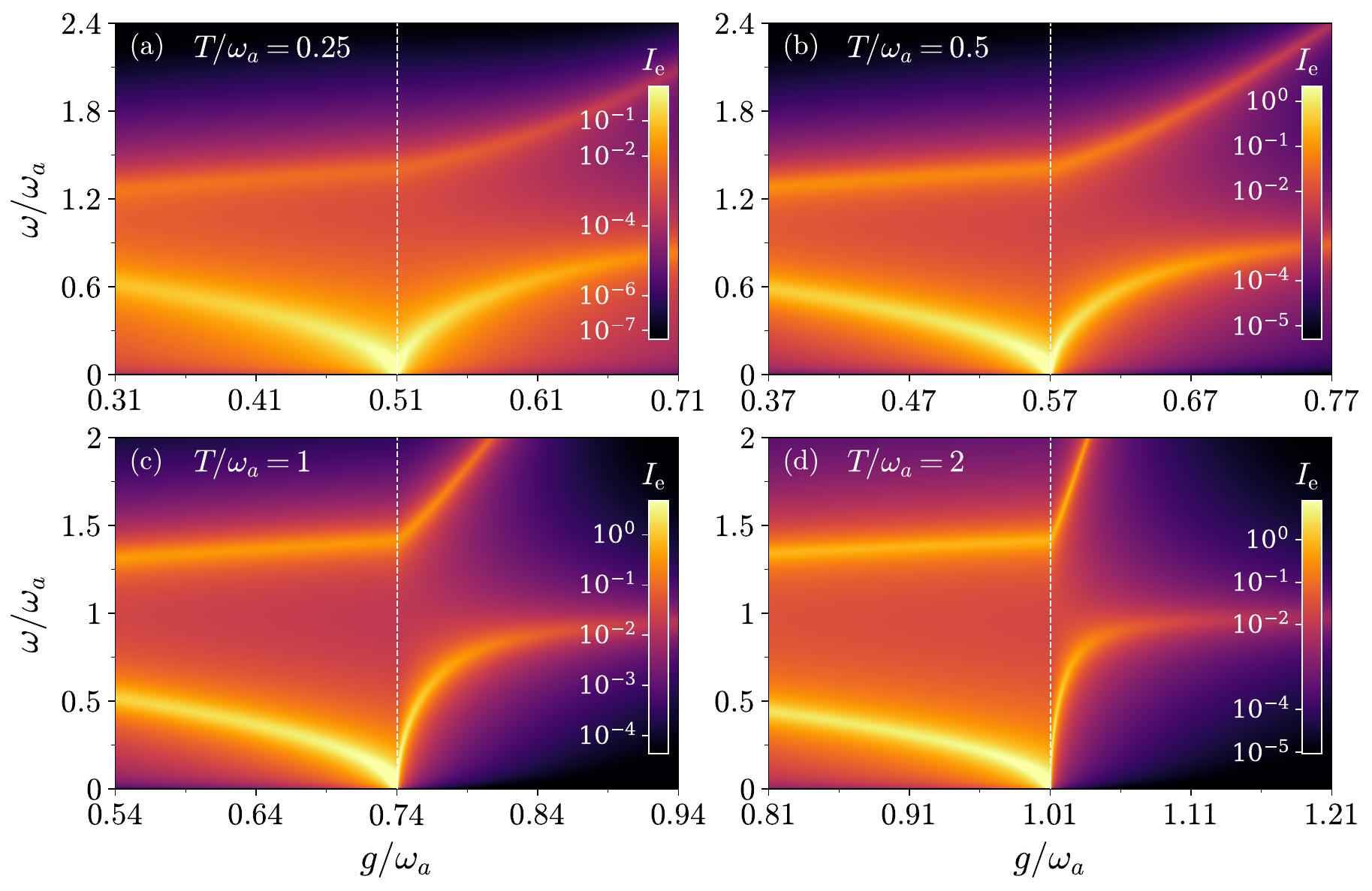}
    \caption{\textbf{Incoherent thermal emission spectra.}
    Logarithmic color maps of $|S_{12}(\omega)|^2 n_B(\omega)$, obtained from the cross-scattering coefficient $S_{12}(\omega)$ [\eqref{eq:S12}] weighted by the Bose-Einstein occupation $n_B(\omega)$, as a function of $g/\omega_a$ and $\omega/\omega_a$ for four temperatures: \textbf{(a)}~$T/\omega_a = 0.25$, \textbf{(b)}~$0.5$, \textbf{(c)}~$1$, and \textbf{(d)}~$2$. The green solid lines trace the corresponding polaritonic dispersions in the absence of reservoir coupling. At higher temperatures the post-critical decoupling and the growing damping mismatch in the superradiant phase suppress the thermal  emission. Parameters as in Fig.~\ref{fig:refl_spectra}.}
    \label{fig:incoh_spectra}
\end{figure*}

The Langevin equations derived above determine the frequency-resolved linear response of the system in the thermodynamic limit. To relate this internal dynamics to reflection and emission measurements, we introduce the incoming and outgoing fields associated with the two reservoirs, following the input--output formulation of Refs.~\cite{GardinerCollett1985,YurkeDenker1984}. The incoming fields are constructed from the free-reservoir solutions with initial conditions at $t_0\to-\infty$,
\begin{equation}
C_{{\rm in},j}(t) =\! \int_0^\infty \!\!d\omega\, \sqrt{\frac{\hbar}{4\pi\omega}} \left[ \tilde c_{{\rm in},j}(\omega)e^{-i\omega t} + \tilde c_{{\rm in},j}^{\dagger}(\omega)e^{i\omega t} \right],
\label{eq:Cin}
\end{equation}
where $j=a,b$ and the continuum operators satisfy standard bosonic commutation relations, $[\tilde c_{{\rm in},j}(\omega), \tilde c_{{\rm in},k}^{\dagger}(\omega')] = \delta_{jk}\delta(\omega-\omega')$. The outgoing fields share the same expansion in terms of $\tilde c_{{\rm out},j}(\omega)$ and are obtained from the free-reservoir solutions specified at $t_1\to+\infty$.

Adopting the Fourier convention introduced previously, \eqref{eq:Cin} yields
\begin{equation}
\tilde C_{{\rm in},j}(\omega) = \sqrt{\frac{\pi\hbar}{\omega}}\, \tilde c_{{\rm in},j}(\omega), \qquad \omega>0,
\label{eq:Cin_positive}
\end{equation}
with an identical relation holding for the outgoing fields. The reservoir noise operators are directly related to the incoming fields by
\begin{equation}
\begin{aligned}
\tilde\xi_a(\omega) &= 2\omega\sqrt{\gamma_a(\omega)}\, \tilde C_{{\rm in},a}(\omega), \\
\tilde\xi_b(\omega) &= 2\omega\sqrt{\gamma_b(\omega)\cos\theta}\, \tilde C_{{\rm in},b}(\omega).
\end{aligned}
\label{eq:noise_input}
\end{equation}
Once this link is established, substituting these relations into \eqref{eq:Fin} naturally yields the full Langevin-force vector directly in terms of the physical input fields
\begin{equation}
\tilde F_{\rm in}(\omega) = i\omega
\begin{pmatrix}
\displaystyle \sqrt{\frac{2\gamma_a(\omega)}{\hbar\omega_a}}\, \tilde C_{{\rm in},a}(\omega) \\[7pt]
\displaystyle -\sqrt{\frac{2\gamma_a(\omega)}{\hbar\omega_a}}\, \tilde C_{{\rm in},a}(\omega) \\[7pt]
\displaystyle -\frac{2\langle J_z'\rangle}{\sqrt N} \sqrt{\frac{2\gamma_b(\omega)\cos\theta}{\hbar\overline\omega_b}}\, \tilde C_{{\rm in},b}(\omega) \\[7pt]
\displaystyle \frac{2\langle J_z'\rangle}{\sqrt N} \sqrt{\frac{2\gamma_b(\omega)\cos\theta}{\hbar\overline\omega_b}}\, \tilde C_{{\rm in},b}(\omega)
\end{pmatrix}.
\label{eq:Fin_input}
\end{equation}
This expression is  valid across both phases, safely recovering the normal-phase limit by setting $\theta=0$.

Solving the reservoir equations with final conditions at $t_1\to+\infty$ gives the corresponding output representation, $-i\omega v(\omega) = -i\left(A+\frac{i}{2}\Gamma(\omega)\right)v(\omega) + \tilde F_{\rm out}(\omega)$, where $\tilde F_{\rm out}$ retains the exact form of \eqref{eq:Fin_input} but with the incoming fields replaced by the outgoing ones. Combining this with \eqref{eq:langevinFourier} and eliminating $v(\omega)$ provides the input-output map
\begin{equation}
\begin{aligned}
\tilde F_{\rm out}(\omega) &= M(\omega;A,-\Gamma)\\
&\quad M(\omega;A,\Gamma)^{-1} \tilde F_{\rm in}(\omega).
\end{aligned}
\label{eq:XioutXiin}
\end{equation}

By expressing the force vectors in terms of the continuum reservoir operators, the outgoing cavity field can be written as a linear combination of the incoming noise from both baths:
\begin{equation}
\tilde c_{{\rm out},a}(\omega) = S_{11}(\omega,g,T)\, \tilde c_{{\rm in},a}(\omega) + S_{12}(\omega,g,T)\, \tilde c_{{\rm in},b}(\omega).
\label{eq:Cout_linear}
\end{equation}
Here, the amplitude $S_{11}$ characterizes the direct reflection in the cavity channel, whereas $S_{12}$ describes the conversion of noise from the matter input channel into an outgoing cavity photon. The frequency dependence of both scattering amplitudes is fundamentally governed by the characteristic polynomial $\zeta(\omega) \equiv \det M(\omega;A,\Gamma)$, which evaluates explicitly to
\begin{equation}
\begin{aligned}
\zeta(\omega) ={}& \left[ \omega^2-\omega_a^2+i\omega\gamma_a(\omega) \right] \left[ \omega^2-\overline\omega_b^{\,2} +i\omega\tilde\gamma_b(\omega) \right] \\
&+ 8\bar g^{\,2}\langle J_z'\rangle \omega_a\overline\omega_b.
\end{aligned}
\label{eq:zeta_explicit_io}
\end{equation}
With this definition, the analytical expressions for the scattering amplitudes read
\begin{equation}
\begin{aligned}
S_{11}(\omega,g,T) &= 1 - \frac{2i\omega\gamma_a(\omega) \left[ \omega^2-\overline\omega_b^{\,2} +i\omega\tilde\gamma_b(\omega) \right]}{\zeta(\omega)}
\\
&=\frac{\zeta(\omega_a,\bar\omega_b,-\gamma_a(\omega),\tilde\gamma_b(\omega))}{\zeta(\omega_a,\bar\omega_b,\gamma_a(\omega),\tilde\gamma_b(\omega))}
\end{aligned}
\label{eq:S11}
\end{equation}
and
\begin{equation}
S_{12}(\omega,g,T) = \frac{8i\omega\bar g\langle J_z'\rangle}{\sqrt N\,\zeta(\omega)} \sqrt{\omega_a\overline\omega_b\, \gamma_a(\omega)\gamma_b(\omega)\cos\theta}.
\label{eq:S12}
\end{equation}

\section{Coherent  and Incoherent Emission Spectra}
\label{sec:spectra}

\begin{figure*}[t!]
    \centering
    \includegraphics[width=\textwidth]{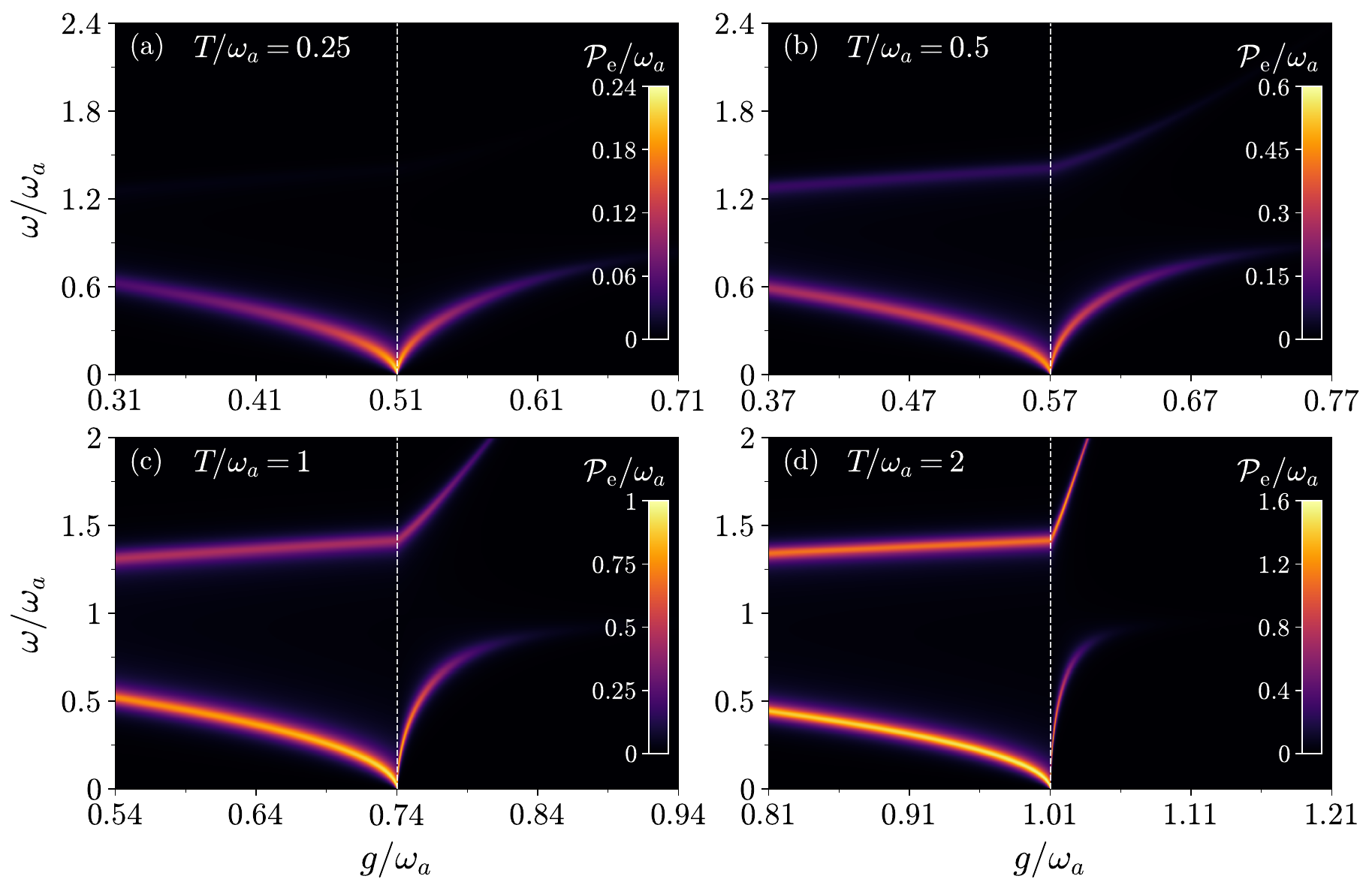}
    \caption{\textbf{Incoherent emitted power spectrum.} 
    Linear color maps of the power spectral density $\mathcal{P}_{\mathrm{e}}(\omega) = \hbar\omega I_{\mathrm{e}}(\omega)$ plotted as a function of the normalized coupling $g/\omega_a$ and frequency $\omega/\omega_a$ around the critical point for four finite temperatures: \textbf{(a)}~$T/\omega_a = 0.25$, \textbf{(b)}~$0.5$, \textbf{(c)}~$1$, and \textbf{(d)}~$2$. The vertical white dashed line in each panel marks the corresponding critical threshold $g_c(T)/\omega_a$. Parameters used: $\omega_a = \omega_b = 1$, $\gamma_a = 0.1$, and $\gamma_b = 0.15$.}
    \label{fig:power_spectra}
\end{figure*}
These scattering amplitudes directly determine the measurable optical properties of the system. They encode the thermodynamic state of the Dicke model through the equilibrium polarization $\langle J_z'\rangle$ and, in the superradiant phase, through the emergence of a non-zero rotation angle $\theta$ and a renormalized atomic frequency $\overline\omega_b$. 

The cavity reflectivity is simply given by $|S_{11}(\omega,g,T)|^2$. The thermal emission spectrum, conversely, follows from the normally ordered correlations of the outgoing field. For independent reservoir inputs with vanishing mean, cross-correlations vanish, and \eqref{eq:Cout_linear} gives the total output photon flux
\begin{equation}
\begin{aligned}
&
\left\langle
\tilde c_{{\rm out},a}^{\dagger}(\omega)
\tilde c_{{\rm out},a}(\omega')
\right\rangle
\\
&=
S_{11}^{*}(\omega,g,T)S_{11}(\omega',g,T)
\left\langle
\tilde c_{{\rm in},a}^{\dagger}(\omega)
\tilde c_{{\rm in},a}(\omega')
\right\rangle
\\
&\quad+
S_{12}^{*}(\omega,g,T)S_{12}(\omega',g,T)
\left\langle
\tilde c_{{\rm in},b}^{\dagger}(\omega)
\tilde c_{{\rm in},b}(\omega')
\right\rangle.
\end{aligned}
\label{eq:flux_conservation}
\end{equation}
To evaluate the incoherent thermal emission generated by the system into the cavity channel, we assume a measurement configuration where the cavity input is in the vacuum state while the matter input is thermally populated at temperature $T$
\begin{equation}
\begin{aligned}
\left\langle \tilde c_{{\rm in},a}^{\dagger}(\omega) \tilde c_{{\rm in},a}(\omega') \right\rangle &= 0, \\
\left\langle \tilde c_{{\rm in},b}^{\dagger}(\omega) \tilde c_{{\rm in},b}(\omega') \right\rangle &= \frac{\delta(\omega-\omega')}{e^{\beta\hbar\omega}-1}.
\end{aligned}
\label{eq:thermal_inputs}
\end{equation}
Under these specific boundary conditions, the incoherent outgoing correlation becomes
\begin{equation}
\begin{aligned}
&\left\langle \tilde c_{{\rm out},a}^{\dagger}(\omega) \tilde c_{{\rm out},a}(\omega') \right\rangle_{\rm incoh}= \delta(\omega-\omega')\, I_{\rm e}(\omega,g,T)\, , \\
&  I_{\rm e}(\omega,g,T)={|S_{12}(\omega,g,T)|^2} n_B(\omega) \,,
\end{aligned}
\label{eq:thermal_emission}
\end{equation}
with $\omega,\omega'>0$ and  $n_B(\omega)=(e^{\beta\hbar\omega}-1)^{-1}$ the Bose-Einstein occupation.
The coefficient of the frequency delta function represents the output occupation spectrum. With the continuum normalization employed here, dividing this quantity by $2\pi$ yields the physical photon flux per unit angular frequency.

We emphasize that this specific configuration serves as an illustrative example of the method's capabilities. The underlying framework is  general and can readily accommodate arbitrary thermal preparations, including non-zero cavity temperatures or thermal imbalances between the reservoirs, thereby offering a highly versatile tool for analyzing non-equilibrium light-matter systems.

To connect the complex-eigenfrequency analysis of the preceding sections with directly measurable input-output observables, we examine the spectroscopic response of the system coupled to ohmic reservoirs, for which the dissipation rates entering the scattering coefficients of \eqref{eq:S11} and \eqref{eq:S12} are frequency-independent, $\gamma_j(\omega)=\gamma_j$ ($j=a,b$). Although a strictly flat spectral density is an idealization over arbitrarily wide bandwidths, our primary focus lies in the low-frequency critical region governed by the softening of the lower polariton, where the constant-damping approximation is well justified; nonetheless, the general input-output expressions derived above can seamlessly accommodate frequency-dependent (sub- and super-ohmic) baths.

We examine two complementary spectroscopic signatures of the resonant Dicke model ($\omega_a=\omega_b=1$): the coherent cavity reflectivity $|S_{11}(\omega)|^2$, mapped in Fig.~\ref{fig:refl_spectra} across six temperatures from $T/\omega_a=0$ to $2$, and the incoherent thermal emission spectrum $I_{\rm e}(\omega)$, shown on a logarithmic scale in Fig.~\ref{fig:incoh_spectra} for four temperatures. In both figures the green solid curves trace the polaritonic dispersions obtained in the absence of reservoir coupling.

A prominent feature emerging from these spectral maps is the presence of a coupling-dependent dissipative frequency shift, analogous to a Lamb shift, which pulls both the reflection minima in Fig.~\ref{fig:refl_spectra} and the emission maxima in Fig.~\ref{fig:incoh_spectra} below the undamped polaritonic eigenfrequencies (green solid lines). However, while this shift is appreciable far from the phase transition, it progressively vanishes in the vicinity of the critical threshold $g_c(T)$, as highlighted in the inset of Fig.~\ref{fig:refl_spectra}(a). At the same rate, the linewidth of the lower-polariton resonance narrows to zero as its frequency minimum approaches $\omega = 0$, ensuring a consistent and well-defined spectral response even in the low-frequency regime.

In the incoherent emission maps of Figs.~\ref{fig:incoh_spectra}(a--d), this critical softening combines with the enhancement of the thermal Bose-Einstein population, $n_B(\omega) \propto T/\omega$ as $\omega \to 0$, which weights the low-energy response and is responsible for the bright emission feature observed around $g_c(T)$.

Comparing the panels of Figs.~\ref{fig:refl_spectra} and \ref{fig:incoh_spectra} shows that temperature affects both the dispersion and the visibility of the polaritonic resonances. As the transition shifts to stronger coupling, the branches separate more sharply beyond threshold, with the lower one flattening near $\omega_a$ and the upper one tracking the renormalized matter frequency [see Figs.~\ref{fig:refl_spectra}(c--f) and \ref{fig:incoh_spectra}(b--d)]. This behavior is consistent with the reduced cavity-matter hybridization discussed in Sec.~\ref{sec:superradiant}. As the modes become predominantly cavity- or matter-like, this reduced hybridization inherently lowers the resonance contrast in both channels. Furthermore, this loss of visibility is enhanced by a growing mismatch between the dissipation rates. Deep in the superradiant phase, the effective atomic damping $\widetilde\gamma_b$ is suppressed by the spin-projection factor $\cos\theta \ll 1$, creating an imbalance relative to the constant cavity decay $\gamma_a$. 
In many experimental implementations, photodetectors directly measure the incoherent power spectrum, so in Fig.~\ref{fig:power_spectra}  we report the normalized emitted power spectral density, defined as $\mathcal{P}_{\mathrm{e}}(\omega)/\omega_a = \hbar\omega I_{\mathrm{e}}(\omega)$. The extra factor of $\omega$ suppresses the low-frequency divergence of the thermal photon occupation, removing the sharp infrared enhancement of the photon flux near criticality and yielding a  regular behavior at $\omega \to 0$. Aside from this regularization, all the physical considerations discussed above regarding the thermal reshaping of the polaritonic dispersions and the superradiant damping mismatch between $\gamma_a$ and $\widetilde{\gamma}_b$ apply equally to $\mathcal{P}_{\mathrm{e}}(\omega)$.

\begin{figure}[!b]
    \centering
    \includegraphics[width=\columnwidth]{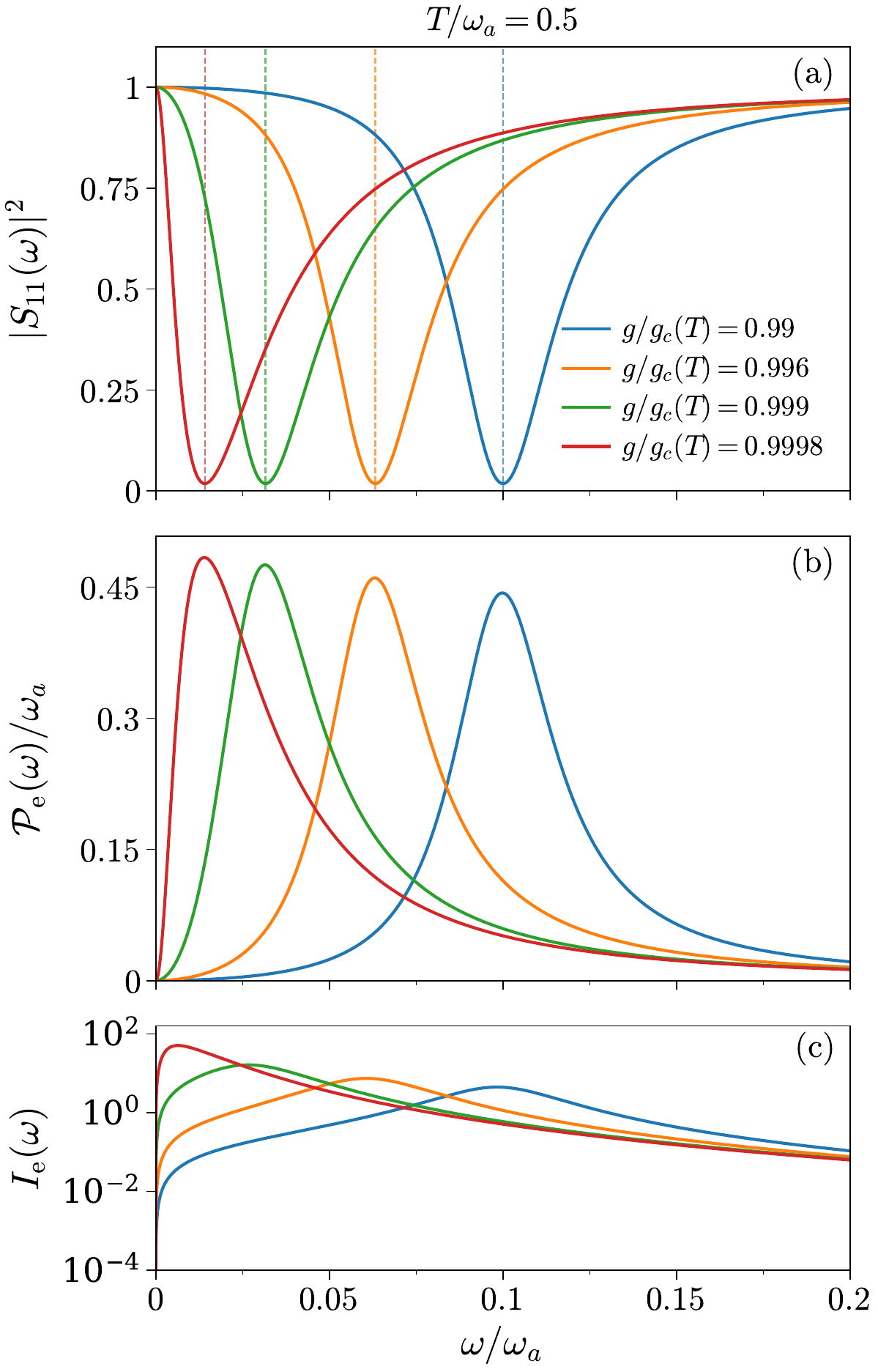}
    \caption{\textbf{Low-frequency spectral lineshapes near the critical point.} 
    Vertical cuts of (a) the reflectivity $|S_{11}|^2$, (b) the normalized emitted power spectrum $P_e/\omega_a$, and (c) the photon flux $I_e$ (logarithmic scale) in the normal phase, at $T/\omega_a=0.5$ and for $g/g_c(T)=0.99,\,0.996,\,0.999,\,0.9998$. In panel~\textbf{(a)}, the vertical dashed lines indicate the corresponding lower-polariton eigenfrequencies in the absence of reservoir coupling. Approaching the critical threshold, the resonances become increasingly asymmetric. Parameters: $\omega_a=\omega_b=1$, $\gamma_a=0.1$, $\gamma_b=0.15$.}
    \label{fig:line_cuts}
\end{figure}

To resolve the spectral lineshapes close to the transition, \figref{fig:line_cuts} shows vertical cuts of the reflectivity $|S_{11}(\omega)|^2$ [Fig.~\ref{fig:line_cuts}(a)], the emitted power spectral density $\mathcal{P}_{\mathrm{e}}(\omega)/\omega_a$ [Fig.~\ref{fig:line_cuts}(b)], and the incoherent photon flux $I_{\mathrm{e}}(\omega)$ [Fig.~\ref{fig:line_cuts}(c)], at $T/\omega_a=0.5$ and for four couplings approaching $g_c(T)$ from the normal phase. The reflection minima and the emission maxima closely follow the undamped lower-polariton eigenfrequencies (vertical dashed lines) rather than the real part of the complex poles [which vanishes in the overdamped region, see Fig.~\ref{fig:disp_rel}]. This confirms that the dissipative frequency shift is negligible in this critical region and explains why experimental fits remain accurate when compared against the  undamped polaritonic dispersions.  As the resonance softens toward $\omega=0$, the lineshapes depart from a symmetric Lorentzian profile becoming increasingly asymmetric. Finally, while the peak photon flux, shown in Fig.~\ref{fig:line_cuts}(c), grows by orders of magnitude as $g\to g_c(T)$ due to the low-frequency thermal enhancement, every emission curve consistently drop to zero as $\omega\to 0$. This regular infrared behavior originates from the vanishing of the dissipative response at zero frequency, a feature naturally captured by the present input-output framework, whereas standard weak-coupling or rotating-wave treatments are not expected to reproduce the low-frequency part of the spectra.

The impedance mismatch discussed above has a counterpart in the opposite limit: when the dissipation rates of the two channels are suitably balanced, reflectionless modes appear, $S_{11}(\omega)=0$, at which radiation entering through the cavity port is entirely absorbed. In our configuration, with the cavity bath at zero temperature and the matter bath thermalized, flux conservation then forces $|S_{12}(\omega)|^2=1$ at that frequency, and the emission collected at the cavity port reaches its maximum value $n_B(\omega)$: a perfect absorber is a perfect emitter, and the system radiates as a black body. Since reflectionless modes require a balance between the dissipation of the two channels, they are naturally connected to $\mathcal{PT}$-symmetric and anti-$\mathcal{PT}$-symmetric non-Hermitian systems ~\cite{Sun2014,Feng2017, Sweeney2020,YangPRL2020,Bonizzoni2026}.
A simple example is the resonant case $\omega_a=\omega_b$ in the normal phase: when the cavity rate matches the effective matter rate, $\gamma_a=\widetilde\gamma_b$, reflectionless modes appear over a broad range of couplings, extending up to the critical point. This property is lost in the superradiant phase, where the renormalized matter frequency $\overline\omega_b$ introduces an effective detuning from $\omega_a$, so that equal damping rates no longer suffice to achieve perfect absorption.

\section{Conclusions}
\label{sec:conclusions}

We have extended the theory of the equilibrium open Dicke model of Ref.~\cite{Lamberto2025} to finite temperature. By formally integrating out the reservoir degrees of freedom, exact quantum Langevin equations valid for arbitrary system-bath coupling strengths are derived, with the ohmic limit being invoked only after establishing the general input-output relations.

The coupling to the reservoirs leaves the temperature-dependent critical curve $g_c(T)$ unchanged for any physical dissipation rate [$\gamma_j(\omega)$ such that $\omega\gamma_j(\omega)\to 0$ as $\omega\to 0$], thus generalizing the zero-temperature result. At any temperature, the lower polariton is overdamped over a finite interval of couplings around $g_c(T)$, and the poles of the reflection amplitude $S_{11}$ coincide with the complex eigenfrequencies of the system. Raising the temperature shifts the transition to stronger couplings, steepens the growth of the order parameter, and accelerates the decoupling of light and matter beyond the critical point. Using the exact input-output theory~\cite{YurkeDenker1984,CollettGardiner1984,GardinerCollett1985,SavastaGirlanda1996}, we calculated reflection and thermal emission spectra across the transition. Away from the critical point, a dissipative frequency shift is visible, which vanishes as the critical point is approached, while the thermal emission becomes brighter around $g_c(T)$ because of the softening of the lower polariton and the growth of the Bose-Einstein occupation at low frequency. We also identified reflectionless modes~\cite{Sun2014,Sweeney2020,Bonizzoni2026}, which appear when the dissipation rates of the two channels are suitably balanced and connect our results with \textit{non-Hermitian} systems~\cite{Feng2017}. At these frequencies the radiation entering through the cavity port is fully absorbed and the emission reaches the blackbody value, so that the system behaves at once as a perfect absorber and a perfect emitter at a given frequency.

 Regarding the ground-state squeezing and entanglement of the ultrastrongly coupled system, we do not expect temperature to alter the conclusions of Ref.~\cite{Lamberto2025}, which demonstrated that such correlations cannot be detected in the static regime. Because these correlations involve virtual excitations that remain bound to the system, thermal noise cannot make them accessible to standard detectors. Conversely, a nonadiabatic modulation of a Hamiltonian parameter can render them observable by converting virtual excitations into real, detectable photons~\cite{Orlando2026}.

The results presented here also suggest some possible applications. In the resonant normal phase, the condition for reflectionless modes, $\gamma_a=\gamma_b\tanh(\beta\hbar\omega_b/2)$, depends on temperature through the atomic polarization. For $\gamma_a<\gamma_b$, a change of temperature could therefore bring the system in and out of perfect absorption, and the resulting reflection dip could provide a thermometric readout of a spin ensemble. Molecular spins coupled to superconducting resonators, in which perfect absorption has been observed~\cite{Bonizzoni2026}, could be a suitable platform. The dependence of the soft mode on temperature near $g_c(T)$ could also be useful for critical thermometry, in line with studies of spin systems and of the dissipative quantum Rabi model~\cite{Aybar2022,Xie2022}. In addition, since $|S_{12}(\omega)|^2$ controls the transfer of energy between the two reservoirs, crossing the transition could be used to tune the heat flow between them~\cite{YamamotoTokura2025,Andolina2024}. Magnonic systems, in which the transition can be tuned by a magnetic field, are natural candidates to test these ideas~\cite{Bamba2022,Kim2025}, although a quantitative assessment of these proposals will require realistic parameters, including the temperature dependence of the intrinsic coupling and loss rates~\cite{Boventer2018}. 
More generally, beyond the polaritonic dispersion, the theoretical framework developed here gives access to the coherent and incoherent spectra of the system in both the normal and superradiant phases, including the position, linewidth, and lineshape of the resonances, thus allowing for a more accurate comparison with finite-temperature experiments. The present approach is not restricted to the Dicke model and can be extended to general light--matter Hamiltonians in the thermodynamic limit [see, \textit{e.g.}, Refs.~\cite{RomanRoche2025,Mercurio2024,parisi2026interplay}].

\begin{acknowledgments}
The authors thank Dr.~Daniele Lamberto for useful discussions. The authors acknowledge the use of Gemini (Google) and Claude (Anthropic) for assistance with language editing, proofreading, and stylistic improvements. All AI-assisted content was reviewed and revised by the authors, who take full responsibility for the final manuscript.
\end{acknowledgments}

\appendix

\section{Thermal spin averages in the effective mean field}

\label{app:spinavg}

At fixed cavity amplitude $\alpha$, the spin part of the mean-field Hamiltonian $H_{\rm MF}(\alpha)$ of \eqref{eq:HMF} is a sum of $N$ identical single-spin Hamiltonians. In the rotated frame defined by \eqrefs{eq:rotation}{eq:theta}, each of them reads $\tfrac12\hbar\overline\omega_b\,\sigma_z'$, where $\sigma_z'$ is the Pauli matrix along the direction of the effective field. As in the finite-temperature Dicke model of Refs.~\cite{HeppLieb1973AnnPhys,WangHioe1973}, the atomic trace in \secref{sec:finiteT} therefore factorizes into $N$ independent two-level traces. Using $(\sigma_z')^2=\mathbf 1$, the single-spin Boltzmann operator expands as

\begin{equation}
\begin{aligned}
e^{-\beta\hbar\overline\omega_b\sigma_z'/2}&
=\cosh\!\left(\frac{\beta\hbar\overline\omega_b}{2}\right)\mathbf 1-\sinh\!\left(\frac{\beta\hbar\overline\omega_b}{2}\right)\sigma_z' ,
\end{aligned}
\end{equation}

which gives the single-atom partition function $2\cosh(\beta\hbar\overline\omega_b/2)$. With $\mathrm{Tr}(\sigma_i')=0$ and $\mathrm{Tr}(\sigma_i'\sigma_j')=2\delta_{ij}$, the thermal averages of the Pauli operators are

\begin{equation}
\begin{aligned}
\langle\sigma_z'\rangle&=-\tanh\!\left(\frac{\beta\hbar\overline\omega_b}{2}\right),
\\
\langle\sigma_x'\rangle&=\langle\sigma_y'\rangle=0 .
\end{aligned}
\end{equation}

Summing over the $N$ identical atoms through $J_i' = \tfrac12\sum_{k=1}^N\sigma_i'^{(k)}$ and rotating back to the original frame, we obtain the collective spin averages

\begin{align}
\langle J_z'\rangle &= -\frac N2\tanh\!\left(\frac{\beta\hbar\overline\omega_b}{2}\right) \, , \notag\\
 \langle J_\pm'\rangle &= 0 \, ,\notag \\
\langle J_x\rangle &= \sin\theta\,\langle J_z'\rangle \, , \notag\\
 \langle J_z\rangle &= \cos\theta\,\langle J_z'\rangle \, ,
\end{align}

which coincide with \eqref{eq:Jz} of the main text and reduce to \eqref{eq:Jz-NP} in the normal phase ($\alpha=0$, $\theta=0$, $\overline\omega_b=\omega_b$). The trace runs over the full $2^N$-dimensional atomic Hilbert space, so that all total-spin sectors are included with their  thermodynamic degeneracies.

\section{Stability of the normal solution and the mean-field free energy}

\label{app:stability}

Here we derive the thermal Hellmann--Feynman identity of \eqref{eq:HF} and the stability condition of the normal phase discussed in \secref{sec:thermaleq}. From $Z_{\rm MF}(\alpha) = \mathrm{Tr}\,e^{-\beta H_{\rm MF}(\alpha)}$, the derivative of the mean-field free energy $F_{\rm MF}(\alpha) = -\beta^{-1}\ln Z_{\rm MF}(\alpha)$ reads

\begin{equation}
\frac{\partial F_{\rm MF}}{\partial\alpha} = -\frac{1}{\beta Z_{\rm MF}(\alpha)}\frac{\partial Z_{\rm MF}}{\partial\alpha} \, .
\label{eq:dFdalpha}
\end{equation}

The operators $H_{\rm MF}(\alpha)$ and $\partial_\alpha H_{\rm MF}(\alpha)$ do not commute, since $[H_{\rm MF},\partial_\alpha H_{\rm MF}] = 4i\hbar^2 g\,\omega_b J_y$. The Duhamel identity for the derivative of the exponential nevertheless holds, and the cyclic property of the trace then merges the two exponentials into $e^{-\beta H_{\rm MF}}$, so that the integral over $s$ contributes a factor $\beta$:

\begin{equation}
\begin{aligned}
\partial_\alpha Z_{\rm MF}
&=-\!\int_0^\beta \!\!\!ds\,\mathrm{Tr}\bigl[e^{-(\beta-s)H_{\rm MF}}\,(\partial_\alpha H_{\rm MF})\,e^{-sH_{\rm MF}}\bigr]\\
&= -\beta\,\mathrm{Tr}\!\left[\frac{\partial H_{\rm MF}}{\partial\alpha}e^{-\beta H_{\rm MF}}\right] \, .
\end{aligned}
\label{eq:dZdalpha}
\end{equation}

Substituting \eqref{eq:dZdalpha} into \eqref{eq:dFdalpha} proves \eqref{eq:HF}. Since $\partial_\alpha H_{\rm MF} = 2N\hbar\omega_a\alpha + 4\hbar g J_x$, the condition $\partial_\alpha F_{\rm MF}=0$ yields the self-consistency condition of \eqrefs{eq:selfcons0}{eq:selfcons}:

\begin{equation}
\begin{aligned}
\frac{\partial_\alpha F_{\rm MF}}{2N\hbar}
&=\omega_a\alpha+\frac{2g}{N}\langle J_x\rangle\\
&=\alpha\Bigl[\omega_a-\frac{4g^2}{\overline\omega_b}
\tanh\!\left(\frac{\beta\hbar\overline\omega_b}{2}\right)\Bigr]=0 .
\end{aligned}
\label{eq:dFMF_explicit}
\end{equation}

Because the spin trace factorizes into $N$ copies of the single-atom partition function of \appref{app:spinavg}, the total partition function is $Z_{\rm MF}(\alpha) = e^{-\beta N\hbar\omega_a\alpha^2}[2\cosh(\beta\hbar\overline\omega_b/2)]^N$, which gives the closed-form mean-field free energy

\begin{equation}
\begin{aligned}
\frac{F_{\rm MF}(\alpha)}{N}&=\hbar\omega_a\alpha^2 -\frac{1}{\beta}\ln\!\left[2\cosh\!\left(\frac{\beta\hbar\overline\omega_b}{2}\right)\right] .
\label{eq:FMFexplicit}
\end{aligned}
\end{equation}

Differentiating \eqref{eq:dFMF_explicit} once with respect to $\alpha$ and evaluating the result at the normal-phase stationary point $\alpha=0$, where $\overline\omega_b=\omega_b$ and $\partial_\alpha\overline\omega_b=0$, gives the free-energy curvature

\begin{equation}
\frac{F_{\rm MF}''(0)}{2N\hbar}=\omega_a-
\frac{4g^2}{\omega_b}\tanh(\beta\hbar\omega_b/2) .
\label{eq:curvature}
\end{equation}

The normal solution $\alpha=0$ is a local minimum of $F_{\rm MF}(\alpha)$ when this curvature is strictly positive, which occurs for $g < g_c(T)$ with $g_c(T)$ defined in \eqref{eq:gcT}~\cite{HeppLieb1973PRA,Hioe1973,Duncan1974}. For $g > g_c(T)$, the origin becomes a local maximum and two degenerate symmetry-broken minima emerge at $\pm\alpha(g,T)$.

\section{Stationary displacements and operator expansion in the superradiant phase}

\label{app:expansion}

In this appendix, we detail the joint system-bath displacement procedure and the large-$N$ fluctuation scaling used in \secref{sec:noneq_excitatoins} and \secref{sec:langevin}. Starting from the unshifted total Hamiltonian $H_{\rm tot}$ in \eqref{eq:Htot_unshifted}, with the shifted-oscillator potential of Refs.~\cite{CaldeiraLeggett1983,Ford1988,Lamberto2025}, we decompose the system and bath coordinates into their expectation values and fluctuations, $X_j = \langle X_j\rangle + \delta X_j$ and $q_{jn} = \langle q_{jn}\rangle + \delta q_{jn}$. The interaction potential of each oscillator then separates into three contributions:

\begin{equation}
\begin{aligned}
&\frac{k_{jn}}{2}(q_{jn}-X_j)^2\\
&=\frac{k_{jn}}2(\langle q_{jn}\rangle-\langle X_j\rangle)^2\\
&+ k_{jn}(\langle q_{jn}\rangle-\langle X_j\rangle)(\delta q_{jn}-\delta X_j) \\
&+ \frac{k_{jn}}{2}(\delta q_{jn}-\delta X_j)^2 \, .
\end{aligned}
\end{equation}

Stationarity with respect to the bath variables requires the terms linear in the bath fluctuations to vanish, which gives $\langle q_{jn}\rangle = \langle X_j\rangle$ and $\langle p_{jn}\rangle = 0$. At these values both the constant energy shift and the linear operator drift vanish identically, leaving only the quadratic fluctuation potential. The reservoirs therefore do not modify the static self-consistency equations of the system. Expressing the bath coordinates through the annihilation operators $c_{jn} = (m_{jn}\omega_{jn}q_{jn} + ip_{jn})/\sqrt{2\hbar m_{jn}\omega_{jn}}$, with $m_{jn}\omega_{jn}^2 = k_{jn}$, the condition $\langle q_{jn}\rangle = \langle X_j\rangle$ gives $\langle c_{jn}\rangle = \sqrt{k_{jn}/(2\hbar\omega_{jn})}\,\langle X_j\rangle$. Inserting the mean values of \eqref{eq:original_system_coordinates}, namely $\langle X_a\rangle = \sqrt{2\hbar N/\omega_a}\,\alpha$ and $\langle X_b\rangle = \sqrt{2\hbar/(N\omega_b)}\sin\theta\,\langle J_z'\rangle$, and writing $c_{an} = c_{s,an} + \sqrt N\overline\sigma_{an}$ and $c_{bn} = c_{s,bn} - \sqrt N\overline\sigma_{bn}$, we recover the induced macroscopic bath amplitudes of \secref{sec:superradiant}, \eqref{eq:sigmabar}.

For the matter bath, we substitute the coordinate fluctuation $\delta X_b = \sqrt{\cos\theta}\,X_{s,b}$ from \eqref{eq:physical_matter_coordinate} and apply the canonical rescaling of \eqref{eq:bath_canonical_transformation}. This rescaling preserves the commutation relations $[q_{s,jn}, p_{s,km}] = i\hbar\delta_{jk}\delta_{nm}$ and transforms each oscillator Hamiltonian as

\begin{equation}
\begin{aligned}
&\frac{\delta p_{bn}^2}{2m_{bn}}
 +\frac{k_{bn}}2(\delta q_{bn}-\delta X_b)^2\\
&\quad=\frac{p_{s,bn}^2}{2M_{bn}}
 +\frac{\tilde k_{s,bn}}2(q_{s,bn}-X_{s,b})^2 ,
\end{aligned}
\end{equation}

with $M_{bn} = m_{bn}\cos\theta$ and $\tilde k_{s,bn} = k_{bn}\cos\theta$, which establishes \eqref{eq:Hint}.

Finally, to verify the $N$-scaling of the spin fluctuation operators in the rotated frame, we evaluate their variances in the factorized thermal reference state of \appref{app:spinavg}. Since the $N$ spins are statistically independent at fixed mean field, the variances of the collective components equal $N$ times the single-spin variances:

\begin{align}
\langle (J_x')^2\rangle &= \langle (J_y')^2\rangle = \frac{N}{4} \, , \notag\\
\langle L^2\rangle &= \frac{\langle (J_z'-\langle J_z'\rangle)^2\rangle}{N}= \frac{1}{4}\,\mathrm{sech}^2\!\left(\frac{\beta\hbar\overline\omega_b}{2}\right) \, .
\end{align}

The transverse spin fluctuations therefore scale as $J_\pm' = \mathcal{O}(\sqrt N)$, whereas the rescaled longitudinal fluctuation $L = (J_z'-\langle J_z'\rangle)/\sqrt N$ is an intensive operator of order $\mathcal{O}(1)$, in accordance with the quantum central limit theorem~\cite{Goderis1989,Goderis1990}, which also underlies the treatment of longitudinal fluctuations in open Dicke models~\cite{Boneberg2022}.

\section{Quantum Langevin equations and input-output relations}

\label{app:E}

This appendix derives the quantum Langevin equations in the presence of the shifted interaction Hamiltonian of \eqref{eq:Hint} and establishes the analytical scattering amplitudes of \secref{sec:io}, generalizing the zero-temperature procedure of Ref.~\cite{Lamberto2025}. Throughout, $H_{\rm sys}+H_{\rm int}$ denotes the shifted total Hamiltonian, in which the system coordinates are $X_{s,j}$ and the bath variables are $(q_{s,jn},p_{s,jn})$, as opposed to the unshifted $H_{\rm tot}$ of \eqref{eq:Htot_unshifted}.

\subsection{Derivation of the quantum Langevin equations}

\label{app:qle_derivation}

From the Hamiltonian $H_{\rm sys}+H_{\rm int}$, the Heisenberg equations of motion for the shifted bath coordinates and momenta read $\dot q_{s,jn} = p_{s,jn}/M_{jn}$ and $\dot p_{s,jn} = -\tilde k_{s,jn}(q_{s,jn}-X_{s,j})$. Combining them gives the driven oscillator equation $\ddot q_{s,jn} + \omega_{jn}^2 q_{s,jn} = (\tilde k_{s,jn}/M_{jn})X_{s,j}(t)$. Integrating this equation from an initial preparation time $t_0$ and integrating the retarded term by parts, the force exerted by the $j$th reservoir on the system, $\sum_n\tilde k_{s,jn}[q_{s,jn}(t)-X_{s,j}(t)]$, takes the form

\begin{equation}
\begin{aligned}
&\xi_j(t)-\sum_n\tilde k_{s,jn}\cos[\omega_{jn}(t-t_0)]\,X_{s,j}(t_0)\\
&\quad-\int_{t_0}^t dt'\sum_n\tilde k_{s,jn}\cos[\omega_{jn}(t-t')]\dot X_{s,j}(t').
\end{aligned}
\label{eq:bath_force_app}
\end{equation}

The bath noise operator $\xi_j(t)$, built from the free evolution of the reservoir modes, is

\begin{equation}
\begin{aligned}
\xi_j(t)&=\!\sum_n\!\sqrt{\hbar\tilde k_{s,jn}\omega_{jn}/2}\,\bigl[c_{s,jn}(t_0)e^{-i\omega_{jn}(t-t_0)}\!+\!\mathrm{H.c.}\bigr].
\label{eq:xi_def_app}
\end{aligned}
\end{equation}

Substituting this solution into the Heisenberg equation of motion for an arbitrary system operator $Y(t)$ gives the general quantum Langevin equation~\cite{Ford1988,GardinerZoller2004}

\begin{equation}
\begin{aligned}
\dot Y(t)={}&\frac{i}{\hbar}[H_{\rm sys},Y(t)]-\!\sum_j\frac{i}{2\hbar}\Bigl\{[X_{s,j},Y(t)],\xi_j(t)\\
&-f_j(t-t_0)X_{s,j}(t_0)-\!\int_{t_0}^t\! \!\!dt'\,f_j(t-t') \dot X_{s,j}(t')\Bigr\}_+,
\end{aligned}
\label{eq:QLE_with_slip}
\end{equation}

where $\{\cdot\,,\cdot\}_+$ denotes the anticommutator. Since $\tilde k_{s,an}=k_{an}$ and $\tilde k_{s,bn}=k_{bn}\cos\theta$, the sums over $n$ in \eqref{eq:bath_force_app} are the reference kernel $f_j(t)=\sum_nk_{jn}\cos(\omega_{jn}t)$ for the cavity, while for the matter the kernel in the equation above is $\cos\theta\,f_b(t)$.  Taking the preparation time to the remote past, $t_0\to-\infty$, the boundary term proportional to $X_{s,j}(t_0)$ drops out, since the kernel decays for a continuum bath with a smooth spectral density, and \eqref{eq:QLE_with_slip} reduces to \eqref{eq:langevin-general} of the main text.

To evaluate \eqref{eq:QLE_with_slip} for the five system fluctuation operators $(\delta a, \delta a^\dagger, J_-', J_+', L)$, we use the canonical commutator $[X_{s,a},\delta a] = -\sqrt{\hbar/(2\omega_a)}$ together with the rotated spin algebra $[J_+',J_-'] = 2J_z'$ and $[L,J_\pm'] = \pm J_\pm'/\sqrt N$. From \eqref{eq:Xs}, the commutators with the shifted matter coordinate read

\begin{align}
[X_{s,b},J_\mp'] &= \pm\sqrt{\frac{2\hbar}{N\overline\omega_b}}\,\bigl(J_z' - \tan\theta\,J_\mp'\bigr) \, ,\notag \\
[X_{s,b},L] &= \frac{1}{N}\sqrt{\frac{\hbar}{2\overline\omega_b}}\bigl(J_-' - J_+'\bigr) \, .
\end{align}

Inserting them into \eqref{eq:QLE_with_slip}, the equation of motion for the longitudinal operator $L$ contains only operator products that are suppressed by $N^{-1/2}$:

\begin{equation}
\begin{aligned}
\dot L &= \frac{i\bar g}{\sqrt N}(\delta a+\delta a^\dagger)(J_-'-J_+') \\
&-\frac{i}{2N\sqrt{2\hbar\overline\omega_b}}
 \Bigl\{\!J_-'\!-\!J_+',\,\xi_b(t)\!-\!\cos\theta\!\!\int_{-\infty}^t\!\!\!\!\!\!\! dt'f_b(t\!-\!t')\dot X_{s,b}(t')\Bigr\}_+.
\end{aligned}
\label{eq:Ldot_app}
\end{equation}
By the power counting of \appref{app:expansion}, $\delta a, \xi_b, X_{s,b} = \mathcal{O}(1)$, $J_\pm' = \mathcal{O}(\sqrt N)$, and $\bar g = g\cos\theta/\sqrt N = \mathcal{O}(N^{-1/2})$. Every term on the right-hand side of \eqref{eq:Ldot_app} is therefore of order $\mathcal{O}(N^{-1/2})$ and vanishes in the thermodynamic limit, so that $\dot L = 0$ and $-i\omega\tilde L(\omega) = 0$ in the frequency domain. This restriction concerns the dynamical response at nonzero frequency only. In the equations for $J_\pm'$, replacing $J_z' = \langle J_z'\rangle + \sqrt N L$ by its extensive thermal average $\langle J_z'\rangle = \mathcal{O}(N)$ is exact to leading order, while the terms proportional to $L$ and to $\tan\theta\,J_\mp'$ are smaller by $\mathcal{O}(N^{-1/2})$ relative to the leading transverse terms. Keeping the leading order and the commutators above, \eqref{eq:QLE_with_slip} gives \eqrefs{eq:dota}{eq:dotJminus} of the main text for $\delta a$ and $J_-'$, while the equations for $\delta a^\dagger$ and $J_+'$ follow by Hermitian conjugation.

We now Fourier transform these equations for $\omega\neq 0$, where $\tilde L(\omega)=0$, with $\tilde Y(\omega)=\int dt\,e^{i\omega t}Y(t)$. In \eqrefs{eq:dota}{eq:dotJminus}, each memory convolution becomes a product with the transform $\hat f_j(\omega)=\int_0^\infty dt\,f_j(t)e^{i\omega t}$ of the causal kernel. We retain its real part, the dissipative rate $\gamma_j(\omega)=\int_0^\infty dt\,f_j(t)\cos(\omega t)$, and neglect the imaginary part. For the ohmic baths used in \secref{sec:spectra} the neglected part vanishes identically, since $f_j(t)=2\gamma_j\delta(t)$ and $\hat f_j(\omega)=\gamma_j$ is real, so that the equations remain exact in that case. Since the reservoirs couple only to the coordinates, the equations of motion give $-i\omega(\delta\tilde a^\dagger + \delta\tilde a) = i\omega_a(\delta\tilde a^\dagger - \delta\tilde a)$ and, to leading order in $N$, $-i\omega(\tilde J_+' + \tilde J_-') = i\overline\omega_b(\tilde J_+' - \tilde J_-')$. Using them inside the dissipative terms reproduces the matrix equation~\eqref{eq:langevinFourier} with $A$ and $\Gamma(\omega)$ given by \eqrefs{eq:Amatrix}{eq:Gammaomega}.

\subsection{Input-output operators and scattering amplitudes}

\label{app:io_derivation}

To link the noise terms $\xi_j(t)$ in \eqref{eq:xi_def_app} to the input fields $C_{{\rm in},j}(t)$ of \eqref{eq:Cin}, we use the continuum normalization of \secref{sec:io}~\cite{YurkeDenker1984,GardinerCollett1985,SavastaGirlanda1996}. We introduce the spectral densities $k_j(\omega)=\sum_nk_{jn}\delta(\omega-\omega_{jn})$, so that $f_j(t)=\int_0^\infty d\omega\,k_j(\omega)\cos(\omega t)$ and $\gamma_j(\omega)=\tfrac\pi2k_j(\omega)$, together with $\tilde k_{s,a}(\omega)=k_a(\omega)$ and $\tilde k_{s,b}(\omega)=k_b(\omega)\cos\theta$. For ohmic baths $k_j(\omega)=2\gamma_j/\pi$ is constant, which gives $f_j(t)=2\gamma_j\delta(t)$ and $\gamma_j(\omega)=\gamma_j$. In the continuum limit,

\begin{equation}
\begin{aligned}
\xi_j(t)&=\int_0^\infty \!\!\!d\omega'\,
\sqrt{\hbar\omega'\tilde k_{s,j}(\omega')/2}\,\bigl[\tilde c_{{\rm in},j}(\omega')e^{-i\omega't}
 +\mathrm{H.c.}\bigr],
\end{aligned}
\label{eq:xi_cont_app}
\end{equation}

where $\tilde c_{{\rm in},j}(\omega') \equiv c_{s,j}(\omega',t_0)e^{i\omega' t_0}$. With $\tilde k_{s,a}(\omega) = 2\gamma_a(\omega)/\pi$ and $\tilde k_{s,b}(\omega) = 2\gamma_b(\omega)\cos\theta/\pi$, the Fourier transform $\tilde\xi_j(\omega) = \int_{-\infty}^\infty dt\,e^{i\omega t}\xi_j(t)$ of \eqref{eq:xi_cont_app} selects the annihilation operator for $\omega > 0$:

\begin{align}
\tilde\xi_a(\omega) &= \sqrt{4\pi\hbar\omega\gamma_a(\omega)}\,\tilde c_{{\rm in},a}(\omega)\notag\\
&= 2\omega\sqrt{\gamma_a(\omega)}\,\tilde C_{{\rm in},a}(\omega)\notag \, , \\
\tilde\xi_b(\omega) &= \sqrt{4\pi\hbar\omega\gamma_b(\omega)\cos\theta}\,\tilde c_{{\rm in},b}(\omega)\notag\\
&= 2\omega\sqrt{\gamma_b(\omega)\cos\theta}\,\tilde C_{{\rm in},b}(\omega) \, ,
\end{align}

where in the second step we used \eqref{eq:Cin_positive}. This proves \eqrefs{eq:noise_input}{eq:Fin_input}.

To extract the scattering amplitudes $S_{11}(\omega,g,T)$ and $S_{12}(\omega,g,T)$ of \eqrefs{eq:S11}{eq:S12}, we use the identity $M(\omega;A,-\Gamma) = M(\omega;A,\Gamma) + i\Gamma(\omega)$ to rewrite the input-output map of \eqref{eq:XioutXiin} as

\begin{equation}
\tilde F_{\rm out}=(I+i\Gamma M^{-1})\tilde F_{\rm in}.
\label{eq:Fout_Fin_expanded}
\end{equation}

Denoting by $[M(\omega;A,\Gamma)^{-1}]_{jk}$ the $2\times 2$ blocks of the inverse matrix, with $j,k=1,2$ labeling the cavity and matter subspaces, and using the relation between $\tilde F_{\rm in/out}(\omega)$ and $\tilde c_{{\rm in/out},j}(\omega)$ from \eqrefs{eq:Cin_positive}{eq:Fin_input}, we project \eqref{eq:Fout_Fin_expanded} onto the cavity output channel by multiplying from the left with the row vector $\tfrac12(1,\,-1,\,0,\,0)$. Since $(1,-1)\Gamma_{aa}=2\gamma_a(1,-1)$, the factor two from the cavity damping block cancels the projector normalization. Suppressing the common frequency arguments, this yields \eqref{eq:Cout_linear} with

\begin{align}
S_{11}&=1+i\gamma_a(1,-1)[M^{-1}]_{11}\binom{1}{-1},\\
S_{12}&=-\frac{2i\langle J_z'\rangle}{\sqrt N}
 \sqrt{\frac{\omega_a\gamma_a\gamma_b\cos\theta}{\overline\omega_b}}\,(1,-1)[M^{-1}]_{12}\binom{1}{-1}.
\end{align}

The two contractions of the adjugate matrix are

\begin{align}
&(1,-1)[M^{-1}]_{11}\binom{1}{-1}=-\frac{2\omega}{\zeta}
 \bigl(\omega^2-\overline\omega_b^2+i\omega\tilde\gamma_b\bigr),\notag\\
&(1,-1)[M^{-1}]_{12}\binom{1}{-1}=-\frac{4\omega\overline\omega_b\bar g}{\zeta}.
\end{align}

Here $\zeta=\det M$ is given by \eqref{eq:zeta_explicit_io}, and substitution recovers \eqrefs{eq:S11}{eq:S12}. For real frequencies $\omega>0$ and real damping rates $\gamma_j(\omega)$, the squared moduli of \eqrefs{eq:S11}{eq:S12} satisfy

\begin{equation}
\begin{aligned}
|\zeta|^2-|\zeta S_{11}|^2=-32\omega^2\omega_a\overline\omega_b\bar g^2  \langle J_z'\rangle\gamma_a\tilde\gamma_b=|\zeta S_{12}|^2 ,
\end{aligned}
\end{equation}

where we used $\tilde\gamma_b(\omega) = -(2\langle J_z'\rangle/N)\gamma_b(\omega)\cos\theta$ from \eqref{eq:gamma_beff}. Dividing by $|\zeta(\omega)|^2$ gives the unitarity relation $|S_{11}(\omega,g,T)|^2 + |S_{12}(\omega,g,T)|^2 = 1$, which guarantees flux conservation between the two ports across the whole phase diagram.

\bibliography{references}
\end{document}